\documentclass[preprint,twocolumn]{aastex63}
\usepackage[utf8]{inputenc}
\newcommand{\scone}{{\small SCONE}}

\begin{document}
\title{Photometric Classification of Early-Time Supernova Lightcurves with \scone}

\author[0000-0003-1899-9791]{Helen Qu}
\affiliation{Department of Physics and Astronomy, University of Pennsylvania, Philadelphia, PA 19104, USA}
\author[0000-0003-2764-7093]{Masao Sako}
\affiliation{Department of Physics and Astronomy, University of Pennsylvania, Philadelphia, PA 19104, USA}

\correspondingauthor{Helen Qu}
\email{helenqu@sas.upenn.edu}

\begin{abstract}
 In this work, we present classification results on early supernova lightcurves from \scone, a photometric classifier that uses convolutional neural networks to categorize supernovae (SNe) by type using lightcurve data. \scone\ is able to identify SN types from lightcurves at any stage, from the night of initial alert to the end of their lifetimes. Simulated LSST SNe lightcurves were truncated at 0, 5, 15, 25, and 50 days after the trigger date and used to train Gaussian processes in wavelength and time space to produce wavelength-time heatmaps. \scone\ uses these heatmaps to perform 6-way classification between SN types Ia, II, Ibc, Ia-91bg, Iax, and SLSN-I. \scone\ is able to perform classification with or without redshift, but we show that incorporating redshift information improves performance at each epoch. \scone\ achieved 75\% overall accuracy at the date of trigger (60\% without redshift), and 89\% accuracy 50 days after trigger (82\% without redshift). \scone\ was also tested on bright subsets of SNe ($r<20$ mag) and produced 91\% accuracy at the date of trigger (83\% without redshift) and 95\% 5 days after trigger (94.7\% without redshift). \scone\ is the first application of convolutional neural networks to the early-time  photometric transient classification problem.  All of the data processing and model code developed for this paper can be found in the \href{https://github.com/helenqu/scone\ }{\scone\ software package} located at github.com/helenqu/scone \citep{helen_qu_2021_5602043}.

\end{abstract}

\section{Introduction}
Observations of transient and supernova phenomena have informed fundamental discoveries about our universe, ranging from its expansion history and current expansion rate \citep{riess, perlmutter, freedman, riess_2019} to the progenitor physics of rare and interesting events \citep{pursiainen, patrick}. In the near future, next generation wide-field sky surveys such as the Vera C. Rubin Observatory Legacy Survey of Space and Time \citep[LSST,][]{lsst} will have the ability to observe larger swaths of sky with higher resolution and certainly uncover even more new and exciting astrophysical phenomena.

These surveys promise to generate ever larger volumes of photometric data at unprecedented rates. However, the availability of spectroscopic resources is not expected to scale nearly as quickly. Thus, the challenge of effectively allocating these limited resources is more important than ever. For type Ia SN cosmology, spectroscopic information is used to minimize contamination in constructing pure and representative samples of SNe Ia to continue to constrain the dark energy equation of state. For supernova physicists, spectra uncover important information about an event's potential progenitor processes \citep{filippenko, perets, federica, sollerman}. Spectra taken near peak brightness of an event are optimal as they include mostly transient information and are not dominated by host galaxy features.

With millions of alerts each night, fast and accurate automatic classification mechanisms will be needed to replace the time-consuming process of manual inspection. More specifically, the ability to perform classification early on in the lifetime of a transient would allow for ample time to take spectra at the peak luminosity of the event or at multiple points over the course of the event's lifetime.

\subsection{Photometric Supernova Classification}
An impressive body of work has emerged over the past decade on photometric classification of supernovae. Since only a small percentage of discovered supernovae have ever been followed up spectroscopically, a reliable photometric classifier is indispensable to the advancement of supernova science. 

The Supernova Photometric Classification Challenge \citep[SNPhotCC,][]{spcc_1,spcc_2} created not only an incentive to invest in photometric SN classification, but also a dataset that would be used to train and evaluate classifiers for years to come. Successful approaches range from empirical template-fitting \citep{sako2008} to making classification decisions based on manually extracted features \citep{richards, karpenka}. The more recent Photometric LSST Astronomical Time-series Classification Challenge \citep[PLAsTiCC,][]{plasticc} diversified the dataset by asking participants to differentiate between 14 different transient and variable object classes, including the 6 common supernova types included in this work.  The top entries made use of feature extraction paired with various machine learning classification methods, such as boosted decision trees and neural networks \citep{plasticc_results}. Ensemble methods, in which the results of multiple classifiers are combined to create the final classification probability, were widely used as well.

Deep learning is a branch of machine learning that seeks to eliminate the necessity of human-designed features, decreasing the computational cost as well as avoiding the introduction of potential biases \citep{charnock_moss, moss, naul, aguirre}. In recent years, many deep learning techniques have been applied to the challenge of photometric SN classification. 

Recurrent neural networks (RNNs) are designed to learn from sequential information, such as time-series data, and have been used with great success on this problem. \cite{charnock_moss} applies a variant of RNNs known as Long Short Term Memory networks \citep[LSTMs,][]{Hochreiter1997LongSM} to achieve impressive performance distinguishing SNIa from core collapse (CC) SNe. \cite{rapid} uses a gated recurrent unit (GRU) RNN architecture to be able to perform real-time and early lightcurve classification. \cite{moller} performs both binary classification and classification by type with full and partial lightcurves using Bayesian RNNs. \cite{superraenn} uses a GRU RNN as an autoencoder to smooth out irregularities in lightcurve data that is then fed into a random forest classifier.

Convolutional neural networks (CNNs), which are used in this work, are a state-of-the-art image recognition architecture \citep{lecun1989backpropagation, lecun1998gradient, alexnet, zeiler2014visualizing}. \cite{pelican} addresses the issue of non-representative training sets by using a CNN as an autoencoder to learn from unlabeled test data. \cite{alerce_stamp} developed an image time-series classifier as part of the ALeRCE alert broker, using a CNN to differentiate between various transient types as well as bogus alerts.

Outside of these traditional models, deep learning is still providing new and creative solutions to the photometric transient classification problem.  Convolutional recurrent neural networks are used to classify a time series of image stamps by \cite{ramanah} to detect gravitationally lensed supernovae. A newer type of deep learning architecture, known as a transformer, achieves a very impressive result when applied to the PLAsTiCC dataset by \cite{transformer}. A variational autoencoder was used by ParSNIP \citep{parsnip} to develop a low-dimensional representation of transient lightcurves that uses redshift-annotated photometric data to perform full lightcurve photometric classification and generate time-varying spectra, among other tasks.

\subsection{Early Photometric Supernova Classification}

Though much progress has been made on the photometric supernova classification problem, most of the solutions tackle classification of full supernova lightcurves retrospectively. However, the earlier an object can be classified, the more opportunities there are for the community to perform follow-up observation. Spectroscopic or photometric follow-up at early stages not only reveals insights into progenitor physics, but can also serve as a benchmark for further observations at later epochs. SN type IIb, for example, exhibit hydrogen features in early spectra that quickly disappear over time \citep{SNIIb}. Shock breakout physics is another use case of follow-up observation. \cite{patrick} was the first to report capturing the complete evolution of a shock cooling lightcurve, a short-lived event preceding peak luminosity that reveals properties of the shock breakout and progenitor star for stripped-envelope supernovae such as the IIb.

Despite the general focus on full lightcurve classification, several notable works have addressed the challenge of early photometric classification. \cite{sullivan} was able to not only differentiate between SNIa and CC SN, but also predict redshift, phase, and lightcurve parameters for SNIa using only two or three epochs of multiband photometry data. \cite{poznanski} also performed binary Ia vs. CC SNe classification, but using a Bayesian template-fitting technique on only single epoch photometry and photometric redshift estimates. {\small{PSNID}} \citep{sako2008, sako2011}, the algorithm that produced the highest overall figure of merit in SNPhotCC, was used by the Sloan Digital Sky Survey \citep{sdss} and the Dark Energy Survey \citep{des} to classify early-time and full supernova lightcurves.

\cite{rapid} is a recent application of deep learning techniques specifically to early-time transient classification. A GRU RNN is trained and tested on a PLAsTiCC-derived dataset of 12 transients, including 7 supernova types, that are labeled at each epoch with ``pre-explosion" prior to the date of explosion and the correct transient type after explosion. Thus, the model is able to produce a classification at each epoch of observation. \cite{moller} has also produced an RNN-based photometric classifier that is capable of classifying partial supernova lightcurves, but primarily achieves good results for Ia vs. CC SN classification. \cite{villar} uses a recurrent variational autoencoder architecture to perform early-time anomaly detection for exotic astrophysical events within the PLAsTiCC dataset, such as active galactic nuclei and superluminous SNe. Finally, LSST alert brokers such as ALeRCE \citep{alerce_lc} specialize in accurate early-time classification of transient alerts.

\subsection{Overview}
Originally introduced in \cite{Qu_2021}, hereafter Q21, as a full lightcurve photometric classification algorithm, \scone\ was able to retrospectively differentiate Ia vs. CC SN with $>99$\% accuracy and categorize SNe into 6 types with $>98$\% accuracy without redshift information. Our approach centers on producing heatmaps from 2-dimensional Gaussian processes fit on each lightcurve in both wavelength and time dimensions. These “flux heatmaps” of each supernova detection, along with
“uncertainty heatmaps” of the Gaussian process uncertainty, constitute the dataset for our model. This preprocessing step smooths over irregular sampling rates between filters, mitigates the effect of flux outliers, and allows the CNN to learn from information in all filters simultaneously.

Section 2 outlines the details of the datasets and models used in this work and we discuss the classifier's performance on the various dataset types in Section 3, including a comparison with existing literature. We state our conclusions and goals for future work in Section 4.

\section{Methods}
\subsection{Simulations}
For this work, \scone\ was trained and tested on a set of LSST deep drilling field (DDF) simulations. The dataset was created with SNANA \citep{snana} using the PLAsTiCC transient class models for supernovae types Ia, II, Ibc, Ia-91bg, Iax, and SLSN \citep{plasticc_data, plasticc-models, SNIa_1, SNIa_2, SNIa_3, SNIax_1, SNII_1, SNII_2, SNII_3, SNII_4, SNIbc_1, SNIbc_2, SNIbc_3, SNIbc_4, SLSN_1, SLSN_2, SLSN_3}. The relative rates and redshift distribution are identical to those of the data produced for the PLAsTiCC challenge. This is the same dataset used to evaluate \scone's categorical classification performance in Q21.  {No cuts on individual low S/N lightcurve points were made, but lightcurves with fewer than two $5\sigma$ detections were removed, as $t_{\rm trigger}$ would be ill-defined in those cases. We note that in observed data, transient light curve samples will contain SNe contaminated by other galactic astrophysical sources, but methods such as \cite{alerce_lc}} are reliably able to distinguish extragalactic and galactic events. Thus, we can assume the feasibility of creating a pure sample of SN lightcurves such as the one used in this work.

\begin{table}
    \centering
    \caption{Training, validation, and test dataset sizes for the $t_{\rm trigger}+N$ datasets.}
    \label{tbl:datasets}
    \begin{tabular}{l c c}
        \hline
        Dataset & Number of Each Type & Total Size\\
        \hline
        Training & 6148 & 36888\\
        Validation & 769 & 4614\\
        Test & 768 & 4608\\
        \hline
        Full & 7685 & 46110\\
    \end{tabular}
\end{table}

\subsection{Trigger Definition}
We define a \textit{detection} as any observation exceeding the 5$\sigma$ signal-to-noise (S/N) threshold. We define the \textit{trigger} as the next detection that occurs at least one night after the first. In this work, the dataset with the least photometric information includes observations up to (and including) the date of trigger. Thus, all SNe in our datasets have at least two epochs of observation. As the date of first detection is also a common choice of trigger date in other transient surveys, the implications of this discrepancy are explored further in Section 3.3. We present results on a dataset where the distinction between these two definitions is small, i.e. $t_{\rm trigger}\leq t_{\rm first\;detection}+5$.

\subsection{Datasets and Heatmap Creation}
\subsubsection{$t_{\mathrm{trigger}}+N$ Datasets}
To evaluate \scone's classification performance on lightcurves at different stages of the supernova lifetime, five sets of heatmaps were created from the simulations described in Section 2.1. All sets of heatmaps take data starting 20 nights prior to the date of trigger ($t_{\mathrm{trigger}}$) and end at $N=$ 0, 5, 15, 25, and 50 days after the date of trigger, respectively. Hereafter, these are collectively referred to as ``$t_{\mathrm{trigger}}+N$ datasets".

\begin{figure*}[ht]
    \figurenum{1}
    \label{fig:same-sn}
    \includegraphics[scale=0.29,trim={3cm 2cm 0cm 0cm}]{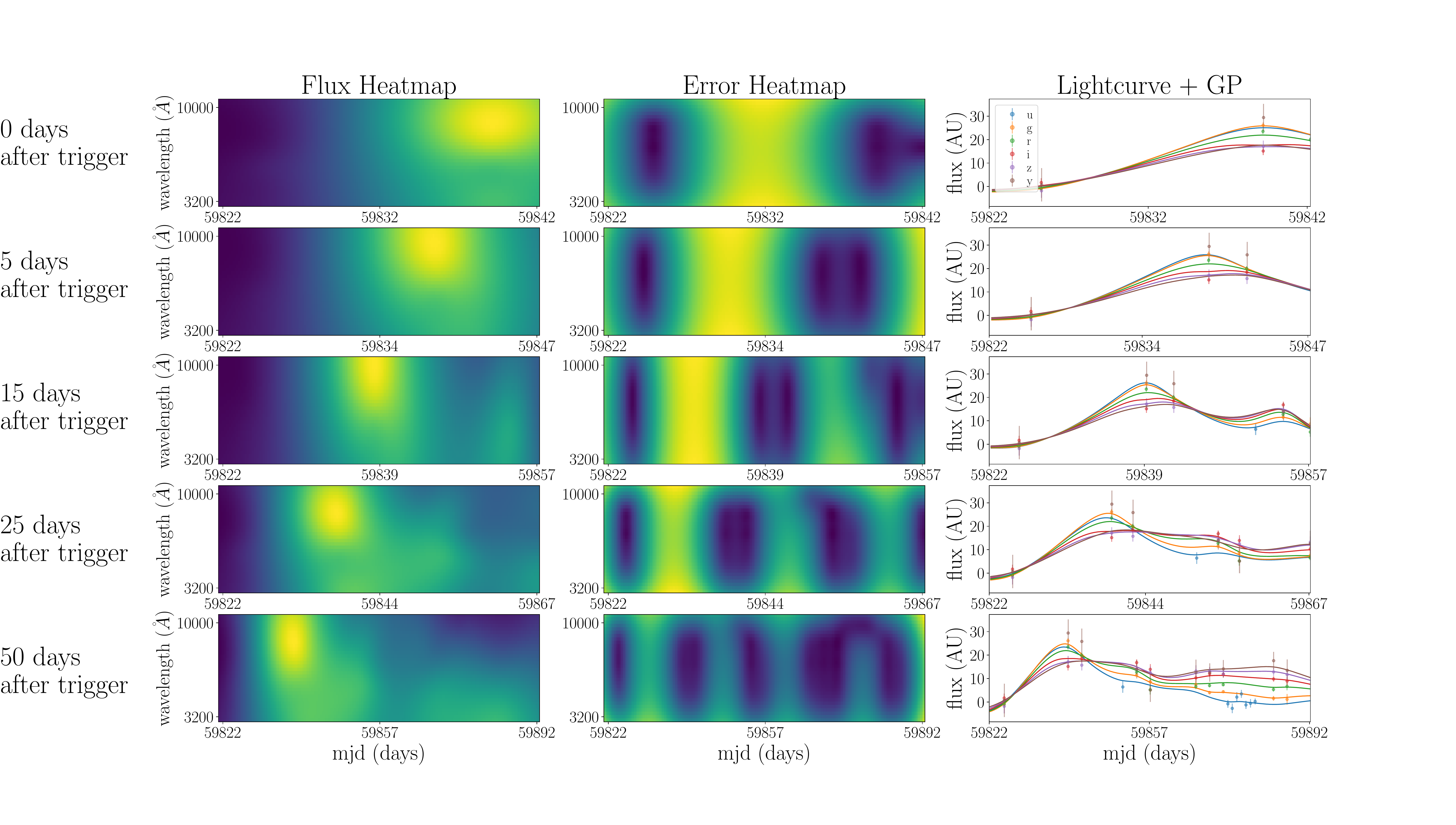}
    \centering
    \caption{An SNII ($z=0.39$) shown in all five heatmap datasets along with the lightcurves and Gaussian process fits used to create each heatmap. The flux and flux error measurements from the raw photometry are shown as points with error bars, while the Gaussian process fits to each photometry band are shown as curves. The Gaussian process errors, which are used to create the heatmaps in the middle column, are not shown in the lightcurve plots. The $x$-axis limit of the plots in each row are different, as the lightcurve is truncated according to the label on the left for each row in the figure. }
\end{figure*}

Prior to training, the lightcurve data is processed into heatmaps. We use the approach described by \citet{avocado} to apply 2-dimensional Gaussian process regression to the raw lightcurve data to model the event in the wavelength ($\lambda$) and time ($t$) dimensions. We use the Mat\'ern kernel ($\nu = \frac{3}{2}$) with a fixed 6000~\AA\  characteristic length scale in $\lambda$ and fit for the length scale in $t$. Once the Gaussian process regression model has been trained, we obtain its predictions on a $\lambda,t$ grid and call this our ``flux heatmap".


It is important to note that the Gaussian processes are fit on lightcurves truncated at $N$ days after trigger in each dataset and not given access to lightcurve information past the cutoff date. Thus, though the $\lambda$ axis is not affected by the different choices of $N$, the $t$ range of the input lightcurve data varies for each $t_{\rm trigger} + N$ dataset. For the datasets in this work, the $\lambda,t$ grids were chosen to preserve the shape of the resulting heatmap despite the fact that the number of nights of lightcurve data varies between the $t_{\rm trigger} + N$ datasets. $\lambda$ is chosen to be $3000 < \lambda < 10,100$~\AA\ with a $221.875$~\AA\ interval for all datasets, while the $t$ interval depends on the number of nights of data: $t_{\rm trigger} - 20 \leq t \leq t_{\rm trigger} + N$ with a $\frac{N+20}{180}$ day interval, where $N=0,5,15,25,50$. This ensures that all heatmaps have size $32 \times 180$.

In addition to the flux heatmap, we also take into account the uncertainties on these predictions at each $\lambda_i, t_j$, producing an ``error heatmap". We stack these two heatmaps depthwise for each SN lightcurve and divide by the maximum flux value to constrain all entries to [0,1]. This $32 \times 180 \times 2$ tensor is our input to the convolutional neural network.

An example of the heatmaps and associated lightcurves of a single SN in all 5 datasets is shown in Figure~\ref{fig:same-sn}. Results on the $t_{\rm trigger}+N$ datasets are described in Section 3.2.

\subsubsection{Bright Supernovae}
Our model was also evaluated on the subset of particularly bright supernovae from the $t_{\rm trigger} +0$ and $t_{\rm trigger} +5$ datasets to emulate a real-world use case of \scone\ for spectroscopic targeting, as bright supernovae are better candidates for spectroscopic follow-up. ``Bright SNe" included in these datasets were chosen to be SNe with last included detection $r<20$ mag. With this threshold, there were 907 SNe in the $t_{\rm trigger} +0$ bright dataset and 5,088 SNe in the $t_{\rm trigger} +5$ bright dataset. As described in more detail in Section 2.4, \scone\ was trained with a standard $t_{\rm trigger}+N$ training set combined with 40\% of the $t_{\rm trigger}+N$ bright dataset, and tested on the $t_{\rm trigger}+N$ bright dataset. Results on these datasets are described in Section 3.5.

\subsubsection{Mixed Dataset}
In order to evaluate \scone's ability to classify SNe with any number of nights of photometry, a sixth dataset (the ``mixed" dataset) was created from the same PLAsTiCC simulations. Data is taken starting 20 nights prior to the date of trigger (as with the $t_{\rm trigger}+N$ datasets) but truncated at a random night between 0 and 50 days after trigger. Due to the choice of the $t$ interval described in Section 2.3.1, heatmaps with any number of nights of photometry data are all the same size can thus be mixed in a single dataset in this manner. We train \scone\ on this mixed dataset and evaluate its performance on each of the $t_{\mathrm{trigger}}+N$ datasets in Section 3.6.

\subsection{Dataset Train/Test Split}
Due to the importance of class balancing in machine learning datasets, the same quantity of SNe from each SN type was selected to create the $t_{\rm trigger} +N$ and mixed datasets used to train, validate, and test \scone. 7685 SNe of each of the 6 types were randomly chosen, as this was the quantity of the least abundant type. Thus, the size of each full dataset was 46,110. An 80/10/10 training/validation/test split was used for all results in this work. The sizes of the training, validation, and test subsets of each dataset can be found in Table \ref{tbl:datasets}.

For evaluation on the bright datasets, \scone\ was trained on a hybrid training set of 40\% of the $t_{\rm trigger} +N$ bright dataset combined with a $t_{\rm trigger} +N$ training set, prepared as described in Section 2.3.1. Thus, the training set was not quite class-balanced, as the bright dataset is not class-balanced but the $t_{\rm trigger}+N$ training set is. The trained model was then evaluated on the full bright dataset to produce the results shown in Figure~\ref{fig:bright}. Due to the imbalanced nature of the bright datasets, the confusion matrices in this figure take the place of an accuracy metric, which could be misleading. We chose to include 40\% of the bright dataset in the training process to ensure that the model has seen enough of these particularly bright objects to make reasonable predictions.

\subsection{Model}
\begin{figure*}
    \figurenum{2}
    \label{fig:architecture}
    \includegraphics[scale=0.50, trim={2.25cm 0cm 0cm 0cm}]{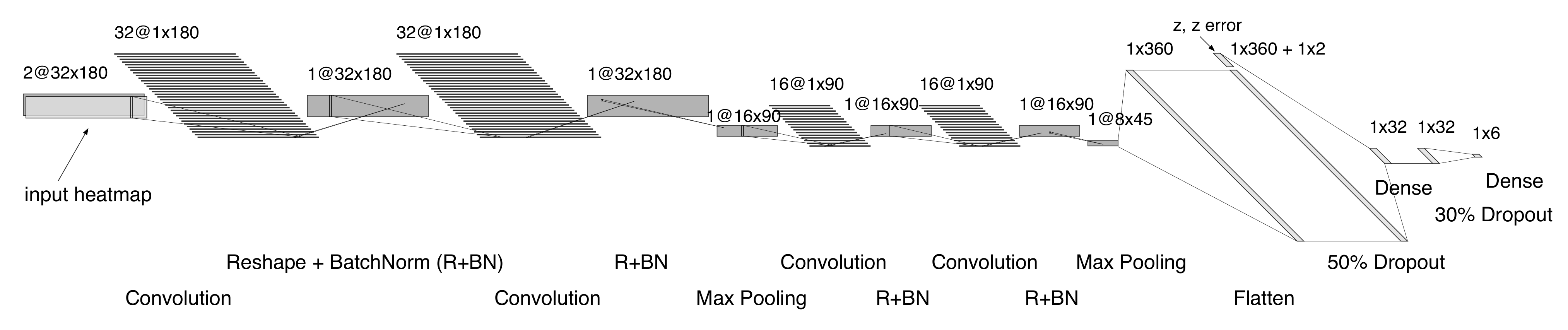}
    \centering
    \caption{\scone\ architecture with redshift information for categorical early lightcurve classification.}
\end{figure*}

In this work, we report early lightcurve classification results using the vanilla \scone\ model developed in Q21 as well as a variant of \scone\ that incorporates redshift information. The architecture of \scone\ with redshift is shown in Figure~\ref{fig:architecture}. Both redshift and redshift error are concatenated with the output of the first dropout layer and used as inputs to the fully connected classifier. The model uses spectroscopic redshift information when available and photometric redshift estimates if not.

Prior to training and testing, the input flux and error heatmaps are divided by the maximum flux value of each heatmap for normalization. This means that absolute brightness information is not used for classification. All results in this work, with and without redshift, used the sparse categorical crossentropy loss function, the Adam optimizer \citep{adam}, and trained for 400 epochs with a batch size of 32. \scone\ without redshift used a constant 1e-3 learning rate, whereas \scone\ with redshift used a constant 5e-4 learning rate.


\subsection{Computational Performance}
The time required for the heatmap creation process was measured using a sample of 100 heatmaps on a single 32-core NERSC Cori Haswell compute node (with Intel Xeon Processor E5-2698 v3). The time required to create one heatmap was $0.03 \pm 0.01$ seconds. When producing larger-scale datasets, this process is also easily parallelizable over multiple cores or nodes to further decrease heatmap creation time.

\scone\ without redshift has 22,606 trainable parameters and \scone\ with redshift has 22,670 trainable parameters, while other photometric classification models require at least hundreds of thousands. The performance gains of this simple but effective model compounded with a small training set make \scone\ lightweight and fast to train. The first training epoch on a NVIDIA V100
Volta GPU takes approximately 17 seconds (4 milliseconds per batch with a batch size of 32), and subsequent training epochs take approximately 5 seconds each with TensorFlow’s dataset caching. The first training epoch on
a Haswell node takes approximately 12 minutes (625 milliseconds per batch), and subsequent epochs take approximately 6 minutes each. Test time per batch of 32 examples is 3 milliseconds on GPU and 10 milliseconds on a Haswell CPU.

\section{Results and Discussion}
\subsection{Evaluation Metrics}
The \textit{accuracy} of a set of predictions describes the frequency with which the predictions match the true labels. In this case, we define our prediction for each SN example as the class with highest probability output by the model, and compare this to the true label to obtain an accuracy.

The \textit{confusion matrix} is a convenient visualization of the correct and missed predictions by class, providing a bit more insight into the model's performance. The confusion matrices shown in Figure~\ref{fig:cm} are normalized such that the $(i,j)$ entry describes the fraction of the true class, $i$, classified as class $j$. The confusion matrices in Figure~\ref{fig:bright} are colored by the normalized values, just like Figure~\ref{fig:cm}, but overlaid with absolute (non-normalized) values. For both figures, the $(i,i)$ entries, or those on the diagonal, describe correct classifications.

The \textit{receiver operating characteristic (ROC) curve} makes use of the output probabilities for each class rather than simply taking the highest probability class, as the previous two metrics have done. We consider an example to be classified as class $i$ if the output probability for class $i$, or $p_i$, exceeds some threshold $p$ ($p_i > p$). The ROC curve sweeps values of $p$ between 0 and 1 and plots the true positive rate (TPR) at each value of $p$ against the false positive rate (FPR). 

TPR is the percentage of correctly classified objects in a particular class, or true positives (TP), as a fraction of all examples in that class, true positives and false negatives (TP+FN). Other names for TPR include \textit{recall} and \textit{efficiency}. The values along the diagonal of the normalized confusion matrices in Figure~\ref{fig:cm} are efficiency values.
$$\mathrm{Efficiency=TPR = \frac{TP}{TP+FN}}$$  
FPR is the percentage of objects incorrectly classified as a particular class, or false positives (FP), as a fraction of all examples not in that class, false positives and true negatives (FP+TN).
$$\mathrm{FPR = \frac{FP}{FP+TN}}$$

The \textit{area under the ROC curve}, or \textit{AUC}, is used to evaluate the classifier from its ROC curve. A perfect classifier would have an AUC of 1, while a random classifier would score (on average) a 0.5.

The \textit{precision} or \textit{purity} of a set of predictions is the percentage of correctly classified objects in a particular predicted class.
$$\mathrm{Precision=\frac{TP}{TP+FP}}$$

\begin{table*}
    \centering
    \caption{Training, validation, and test accuracies \textit{without} redshift information for each early lightcurve dataset. These averages and standard deviations were computed from 5 independent runs of \scone.}
    \label{tbl:no-z-acc}
    \begin{tabular}{l c c c c c}
        \hline
        Accuracy without Redshift & \multicolumn{5}{c}{Days After Trigger}\\
        \hline
        & 0 days & 5 days & 15 days & 25 days & 50 days \\
        \hline
        Training & $58.36 \pm 0.14$\% & $68.92 \pm 0.21$\% & $73.99 \pm 0.14\%$& $76.89 \pm 0.29$\% & $80.93 \pm 0.14$\%\\
        Validation & $59.57 \pm 0.51$\% & $70.74 \pm 0.59$\% & $73.31 \pm 3.01\%$ & $79 \pm 0.84$\% & $82.5 \pm 2.35$\%\\
        Test & $59.66 \pm 0.43$\% & $70.05 \pm 0.63$\% & $73.66 \pm 2.36$\%& $79 \pm 0.86$\% & $82.2 \pm 1.8$\%\\
    \end{tabular}
\end{table*}

\begin{table*}
    \centering
    \caption{Training, validation, and test accuracies \textit{with} redshift information for each early lightcurve dataset. These averages and standard deviations were computed from 5 independent runs of \scone.}
    \label{tbl:with-z-acc}
    \begin{tabular}{l c c c c c}
        \hline
        Accuracy with Redshift & \multicolumn{5}{c}{Days After Trigger}\\
        \hline
        & 0 days & 5 days & 15 days & 25 days & 50 days \\
        \hline
        Training & $72.73 \pm 0.27$\% & $79.61 \pm 0.3$\% & $83.07 \pm 0.2\%$& $84.68 \pm 0.2$\% & $87.17 \pm 0.26$\%\\
        Validation & $74.78 \pm 0.18$\% & $80.52 \pm 1.42$\% & $83.98 \pm 1.15\%$& $86.75 \pm 0.5$\% & $89.2 \pm 0.85$\%\\
        Test & $74.27 \pm 0.51$\% & $80.2 \pm 0.93$\% & $84.14 \pm 1.37$\%& $86.71 \pm 1$\% & $89.04 \pm 0.39$\%\\
    \end{tabular}
\end{table*}

\begin{figure*}[ht]
    \figurenum{3}
    \label{fig:accs}
    \includegraphics[scale=0.4,trim={0 0 0 0}]{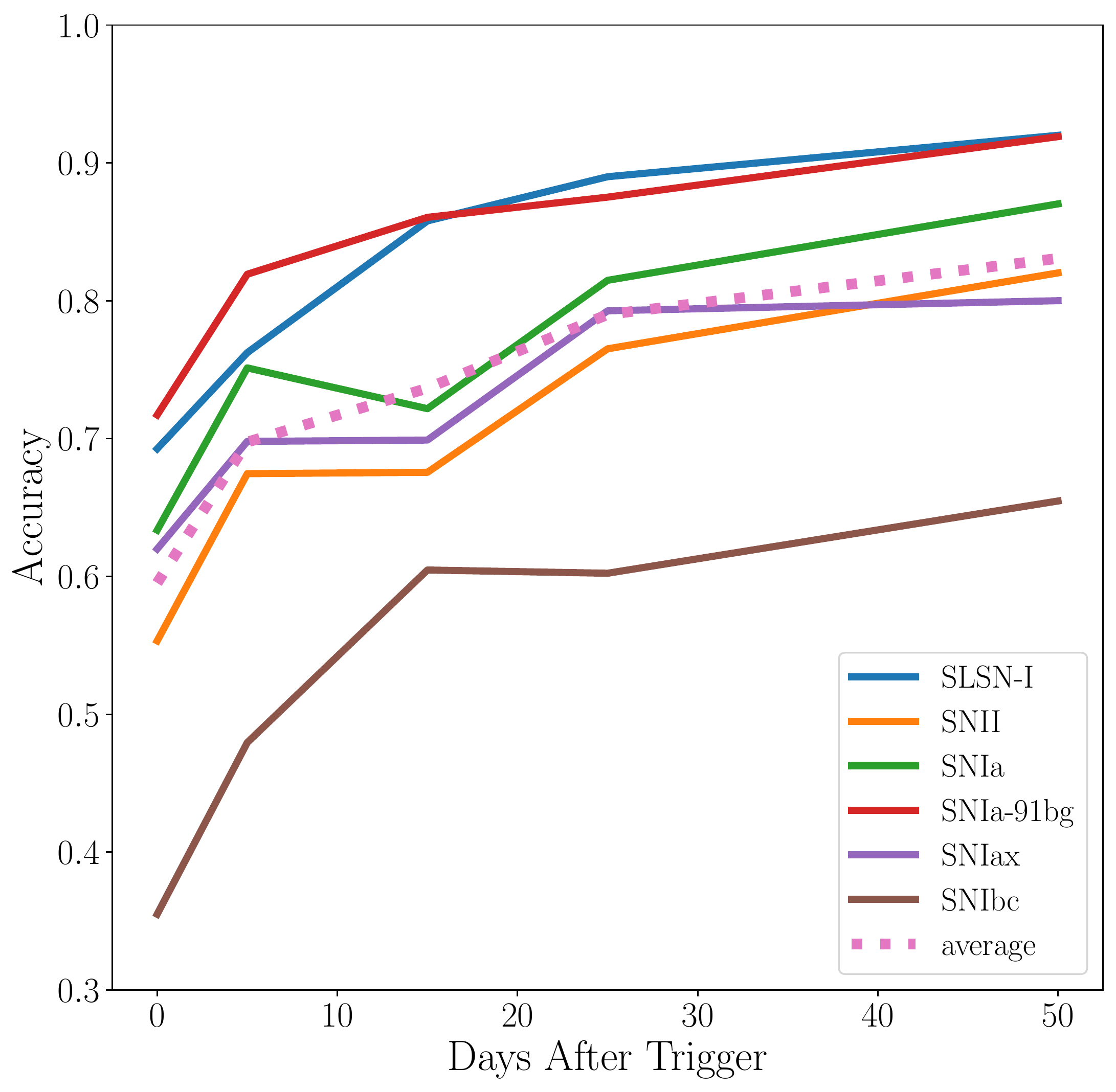}
    \includegraphics[scale=0.4,trim={0 0 0 0}]{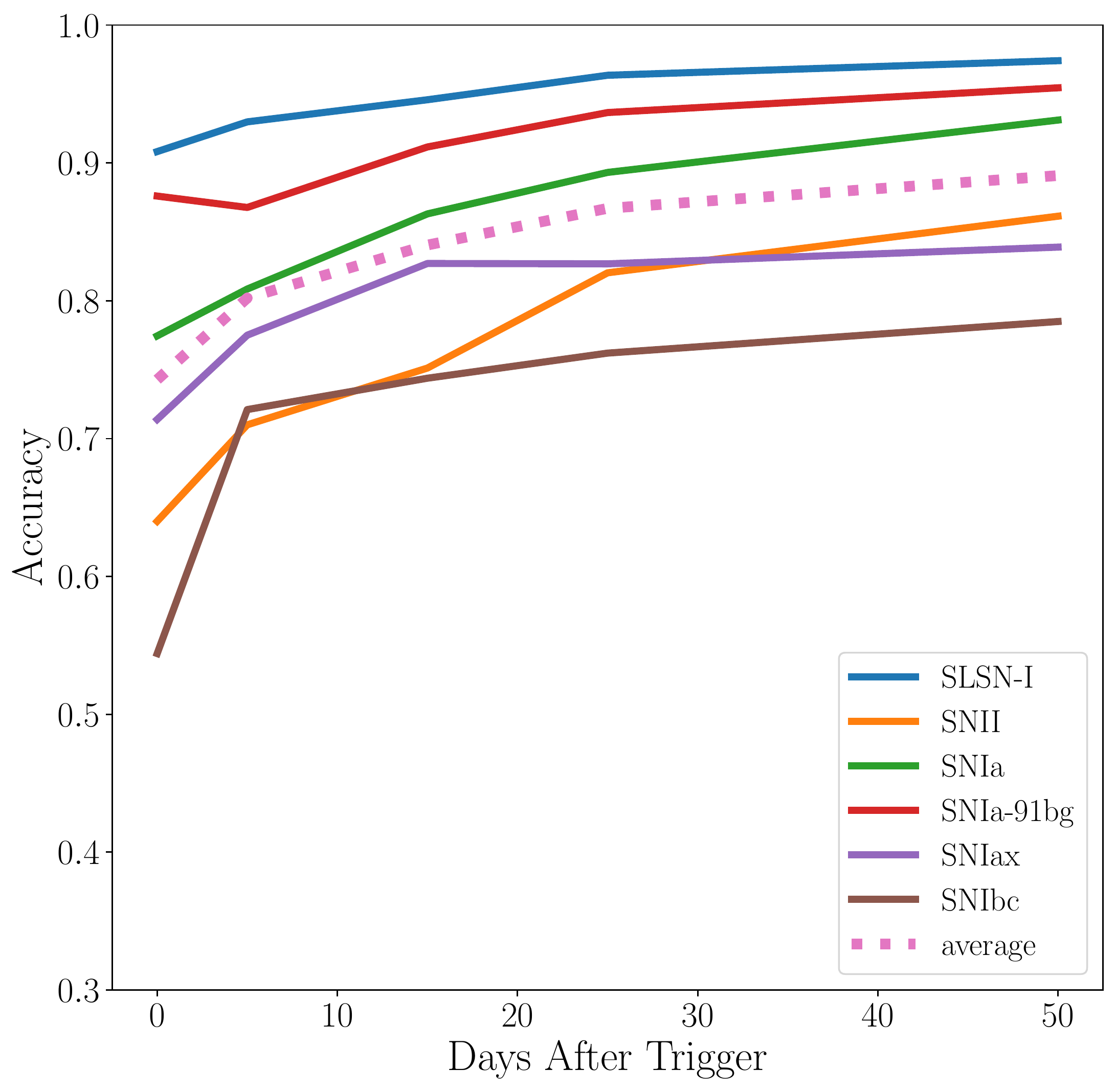}

    \centering
    \caption{Accuracy/efficiency over time for each supernova type without redshift (left) and with redshift (right) for the $t_{\mathrm{trigger}}+N$ test datasets. The values used in this plot correspond with the diagonals on each normalized confusion matrix in Figure~\ref{fig:cm}.}
\end{figure*}

\subsection{$t_{\mathrm{trigger}}+N$ Datasets}
The accuracies our model achieved without redshift on each $t_{\mathrm{trigger}}+N$ dataset are described in Table~\ref{tbl:no-z-acc}, and the accuracies with redshift are described in Table~\ref{tbl:with-z-acc}. These tables show that redshift unequivocally improves classification performance, especially at early times when there is little photometric data to learn from. The inclusion of redshift information not only increases the average accuracies for each dataset but also improves the model's generalizability, as the standard deviations for the validation and test accuracies are lower overall in Table~\ref{tbl:with-z-acc}.

The largest improvement in accuracy between $t_{\mathrm{trigger}}+N$ datasets occurred between 0 and 5 days after trigger for all datasets. Since the explosion likely reached peak brightness during this period, the lightcurves truncated at 5 days after trigger includes much more information necessary for differentiating between the SN types.

Figure~\ref{fig:accs} shows the accuracy evolution over time for each supernova type in the test sets. From the test sets with redshift plot on the right, it is clear that the jump in overall accuracy between 0 and 5 days after trigger can be attributed to the sharp accuracy boost experienced by SNIbc at 5 days after trigger. Overall, SNIbc benefited the most from the inclusion of redshift, though classification performance on all types saw improvement. Note that, as described in Section 2.3, all heatmaps are normalized to values between 0 and 1 so absolute flux values are not used to differentiate between types. Thus, the model cannot rely on relative luminosity information.

\begin{figure*}
    \figurenum{4}
    \label{fig:cm}
    \includegraphics[scale=0.4,trim={1cm 0 0 0}]{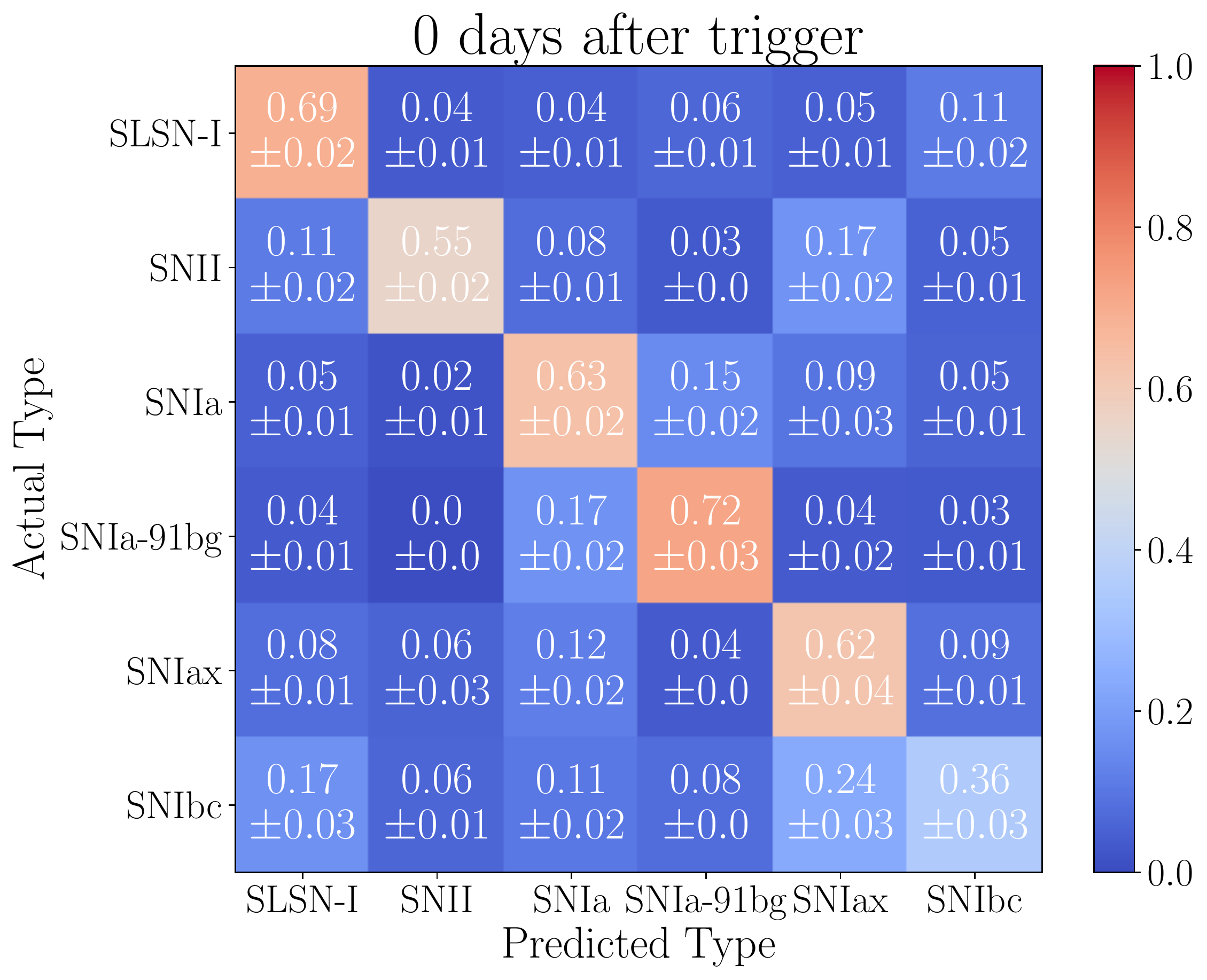}
    \includegraphics[scale=0.4,trim={1cm 0 0 0}]{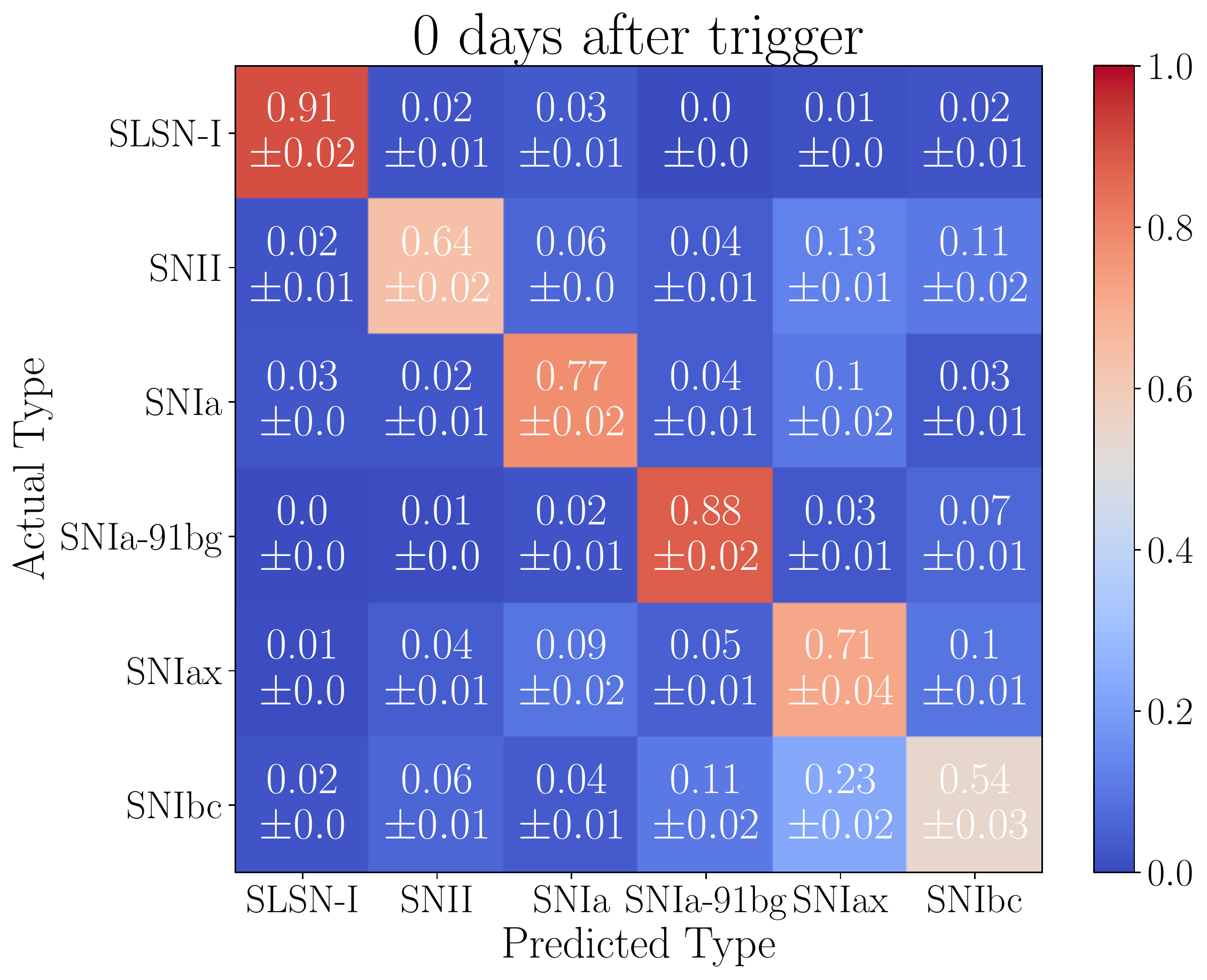}
    \includegraphics[scale=0.4,trim={1cm 0 0 0}]{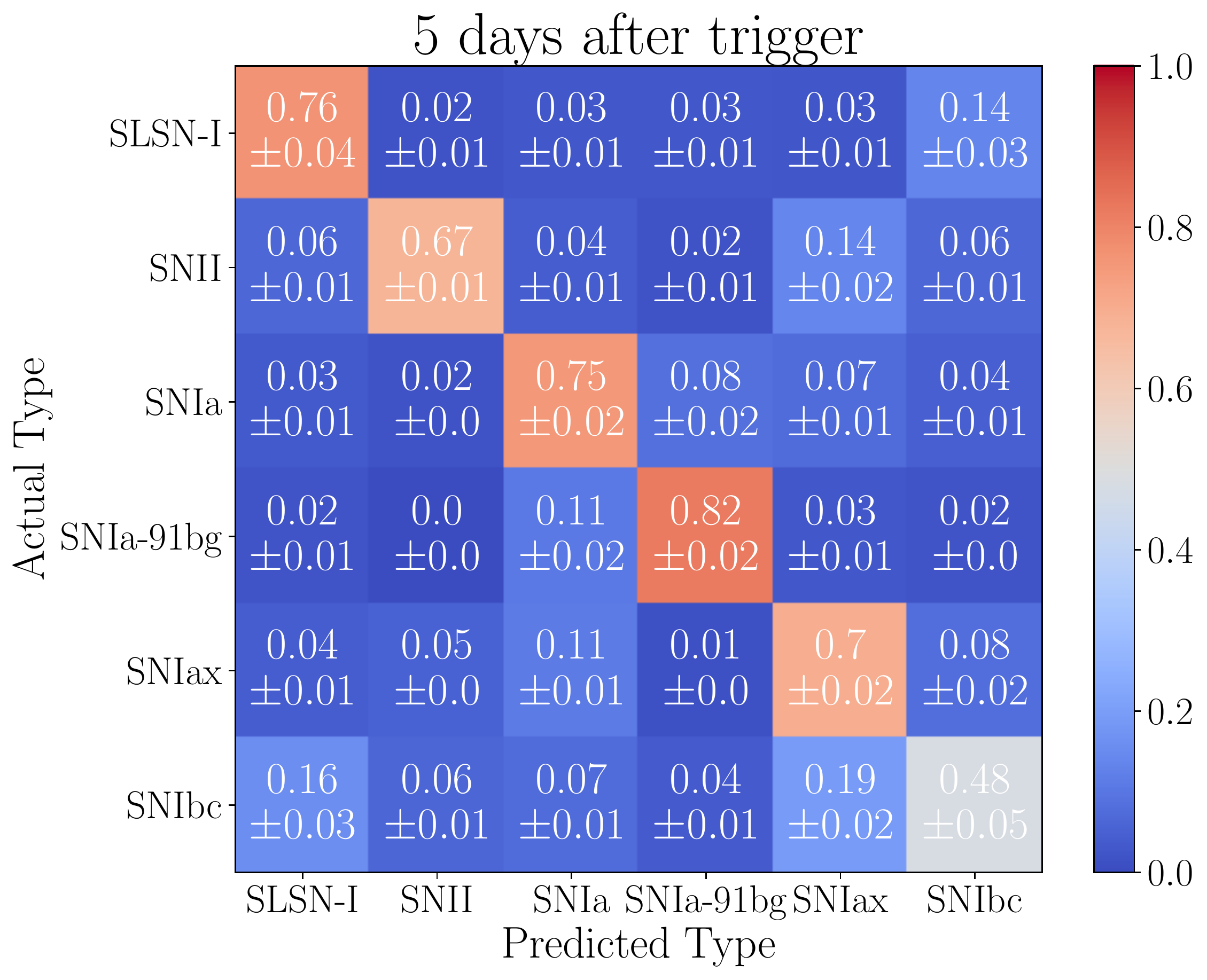}
    \includegraphics[scale=0.4,trim={1cm 0 0 0}]{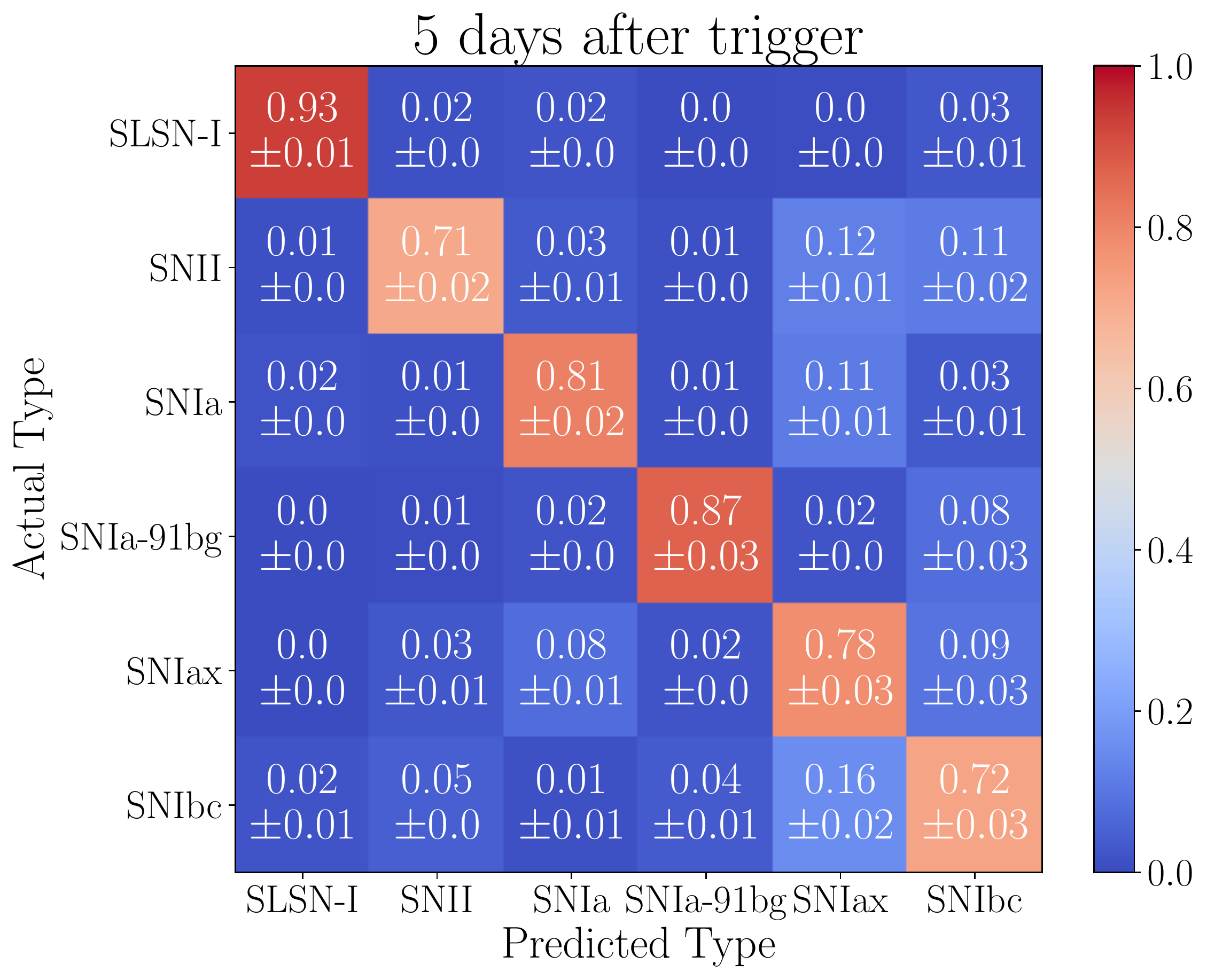}
    \includegraphics[scale=0.4,trim={1cm 0 0 0}]{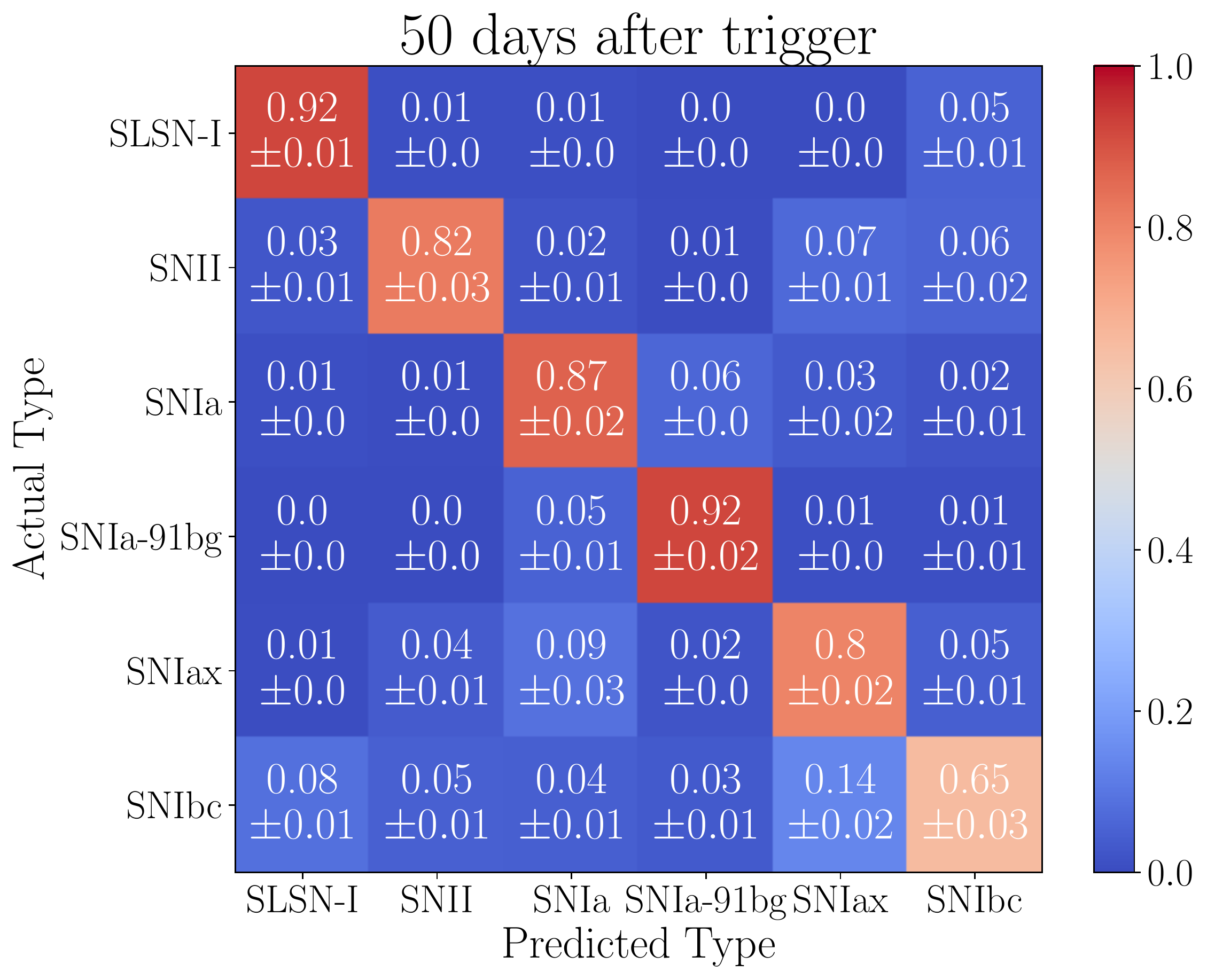}
    \includegraphics[scale=0.4,trim={0cm 0 0 0}]{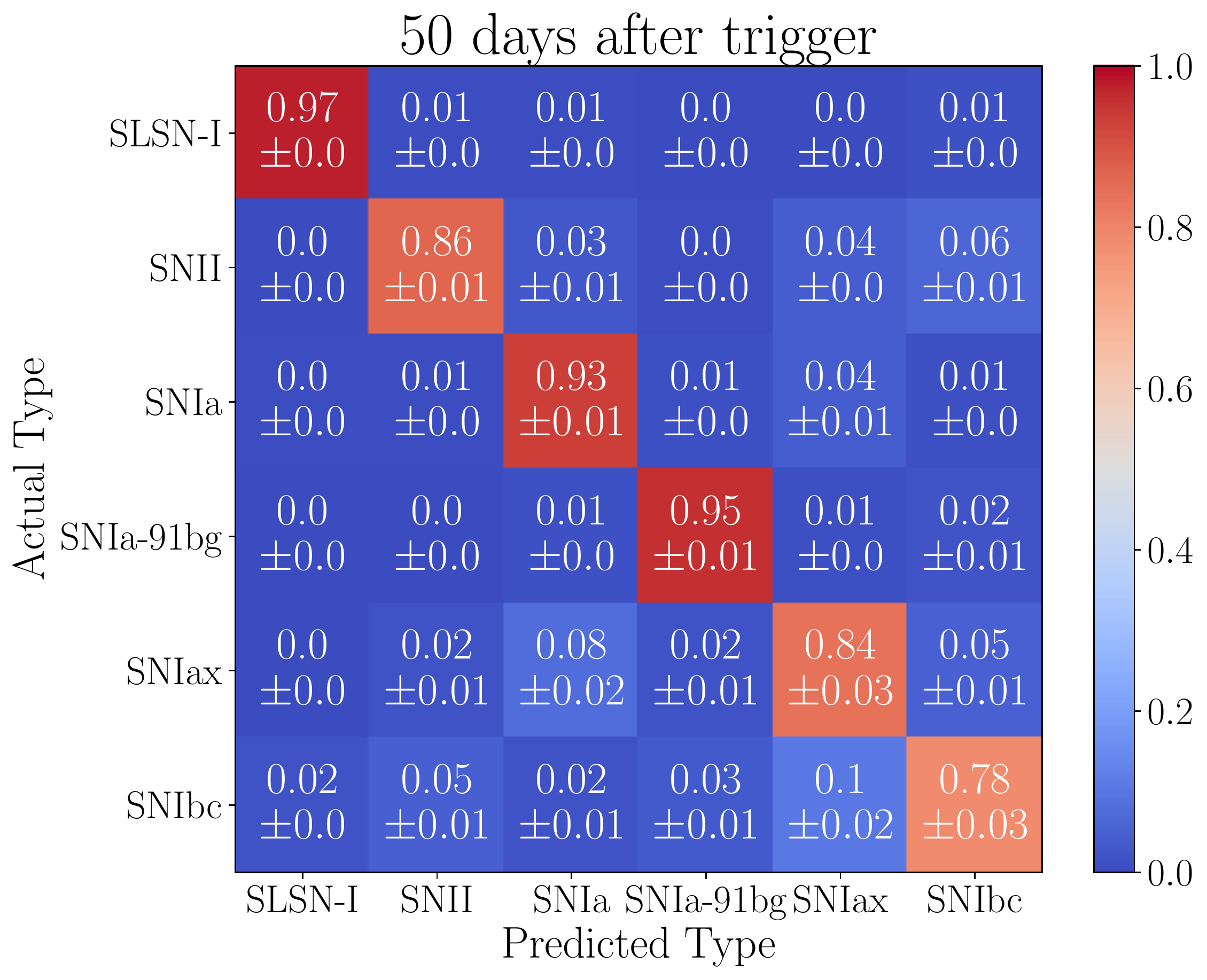}

    \centering
    \caption{Normalized confusion matrices produced by \scone\ without (left) and with (right) redshift for the $t_{\mathrm{trigger}}+\{0,5,50\}$ test sets (heatmaps created from lightcurves truncated at 0, 5, and 50 days after the date of trigger). These matrices were made with test set classification performance from 5 independent runs of \scone.}
\end{figure*}

The confusion matrices for $t_{\mathrm{trigger}}+\{0,5,50\}$ test sets with and without redshift information are shown in Figure~\ref{fig:cm}. The top two panels are early epoch classification results (0 and 5 days after trigger) and the bottom panel shows late epoch results. The confusion matrices from intermediate epochs (15 and 25 days after trigger) were omitted for brevity. 

At the date of trigger (top panel of Figure~\ref{fig:cm}), the incorporation of redshift information primarily prevents confusion between SLSN-I and SNIbc. True SLSN-I events misclassified as SNIbc decreased from 11\% on average to 2\% with redshift. True SNIbc misclassified as SLSN-I decreased from 16\% on average to 2\% with redshift. Overall, SLSN-I were classified with 91\% accuracy with redshift compared to 69\% without redshift, and SNIbc were classified with 54\% accuracy with redshift compared to 36\% without redshift. All types saw marked improvement in classification performance without redshift from 0 to 5 days after trigger, while classification with redshift saw drastic improvement in SNIbc accuracy but only minor improvement for other types. Finally, the effect of added redshift becomes less noticeable by late epochs, where classification accuracy (along the diagonal) is only mildly improved in the bottom panel of Figure~\ref{fig:cm}.

The confusion matrices in Figure~\ref{fig:cm} are normalized by true type, meaning that the values in each row sum to 1. Thus, the values along the diagonal are \textit{efficiency} scores. Normalizing by predicted type, such that the values in each column sum to 1, would result in \textit{purity} scores along the diagonal. However, since all datasets used in Figure~\ref{fig:cm} are class-balanced, the purity scores can be reconstructed from these confusion matrices by dividing each main diagonal value by the sum of the values in its column.

\begin{figure*}
    \figurenum{5}
    \label{fig:roc}
    \includegraphics[scale=0.35,trim={0 0 0 1cm}]{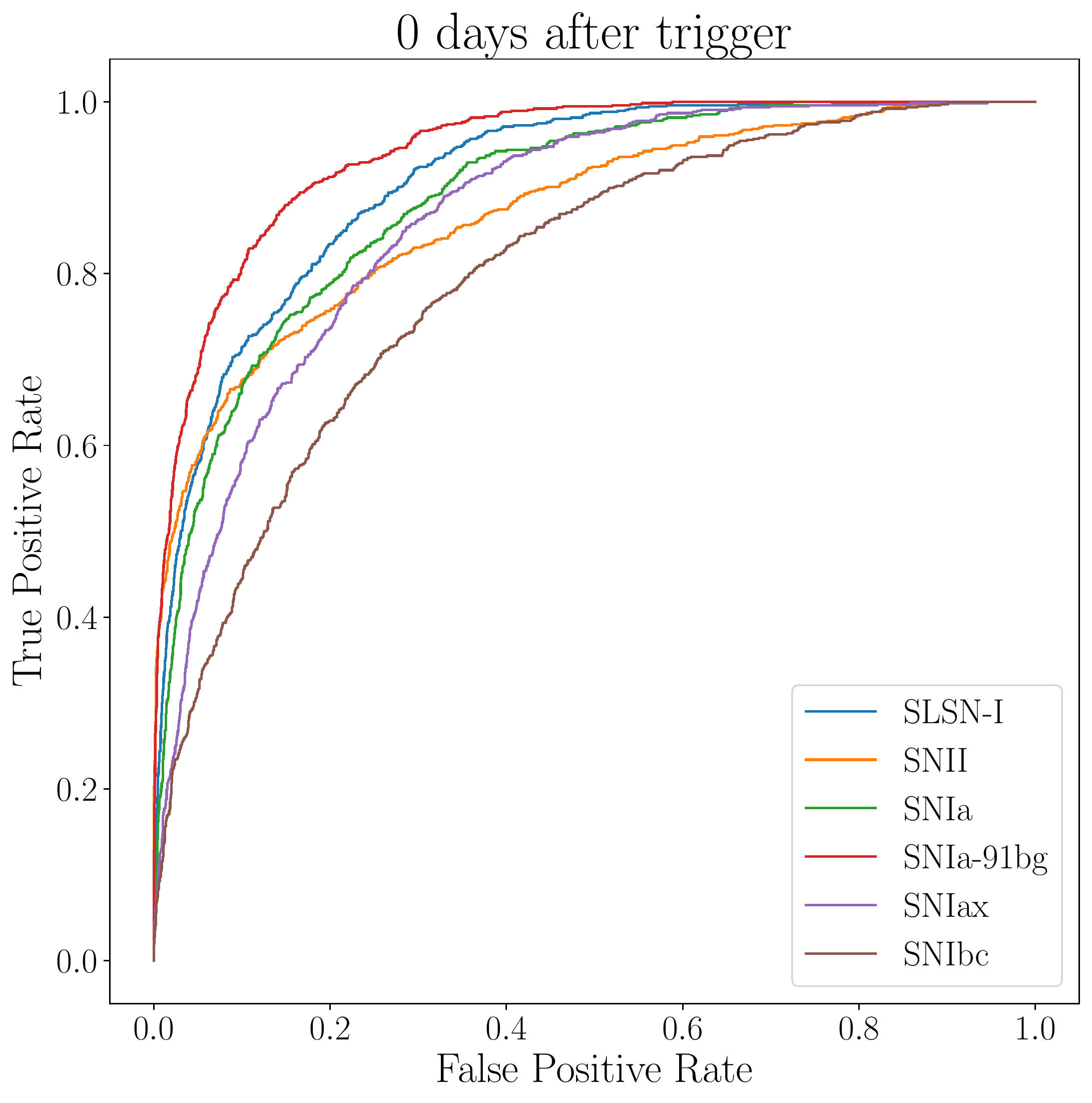}
    \includegraphics[scale=0.35,trim={0 0 0 1cm}]{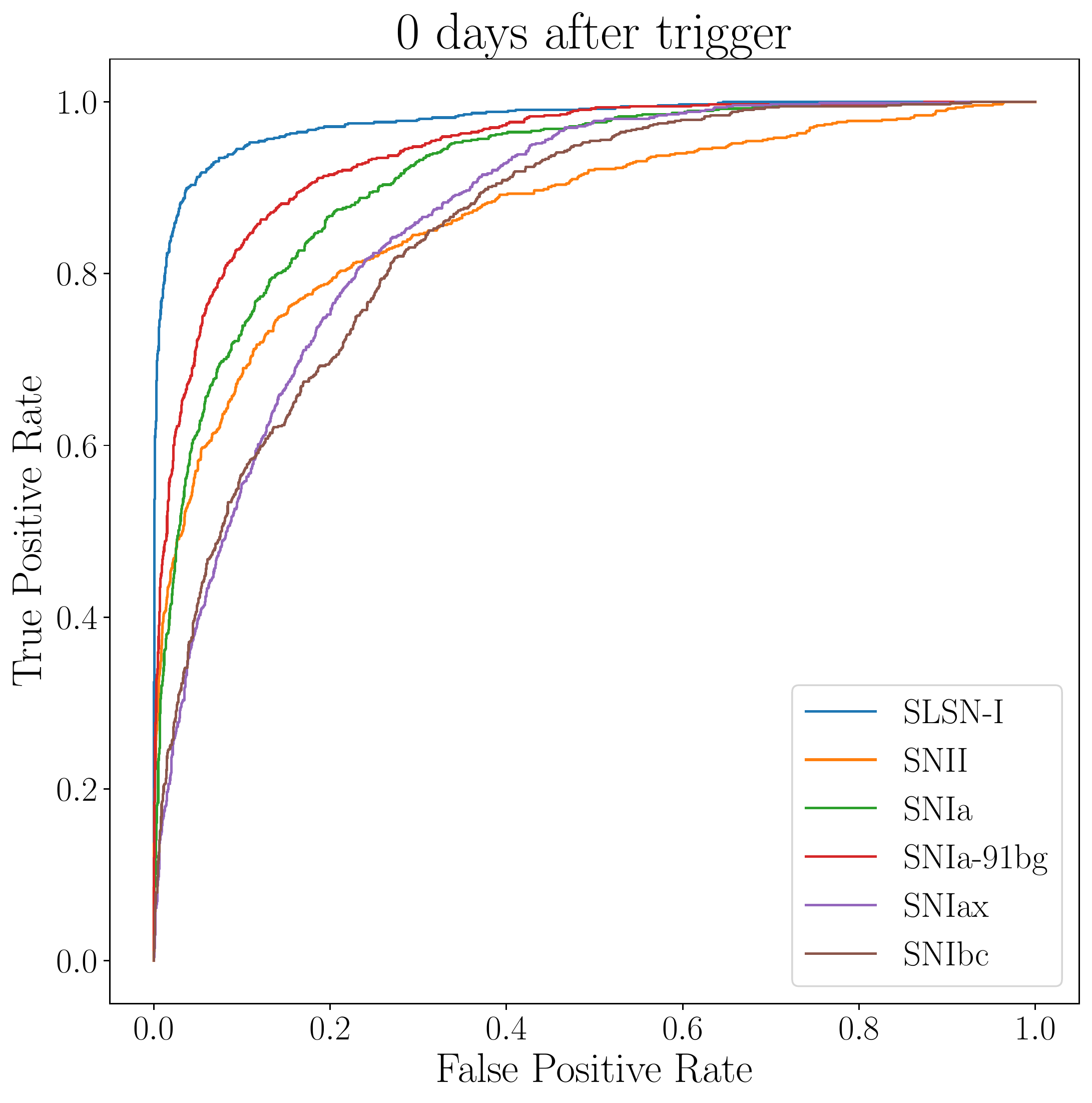}
    \includegraphics[scale=0.35,trim={0 0 0 0}]{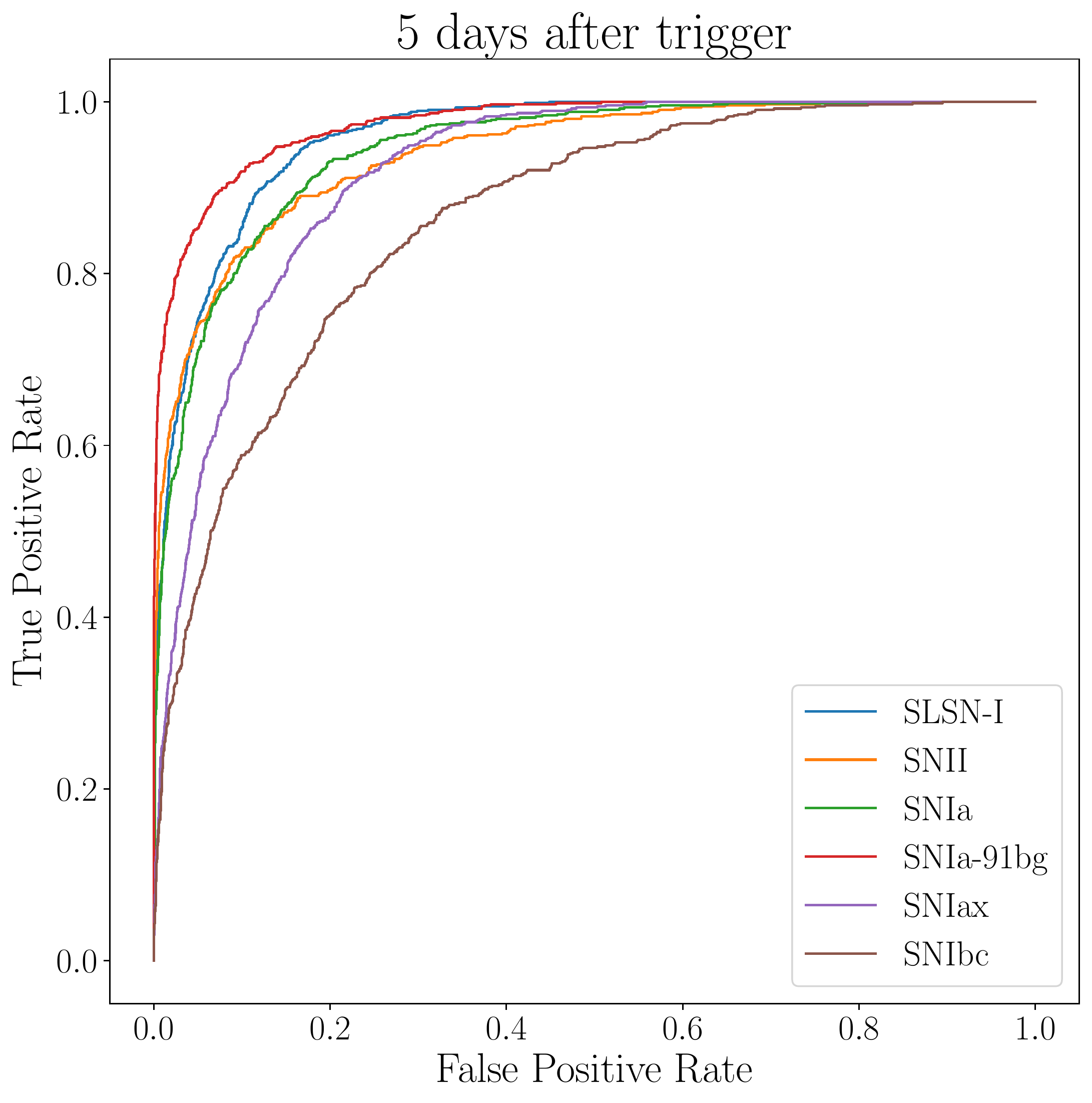}
    \includegraphics[scale=0.35,trim={0 0 0 0}]{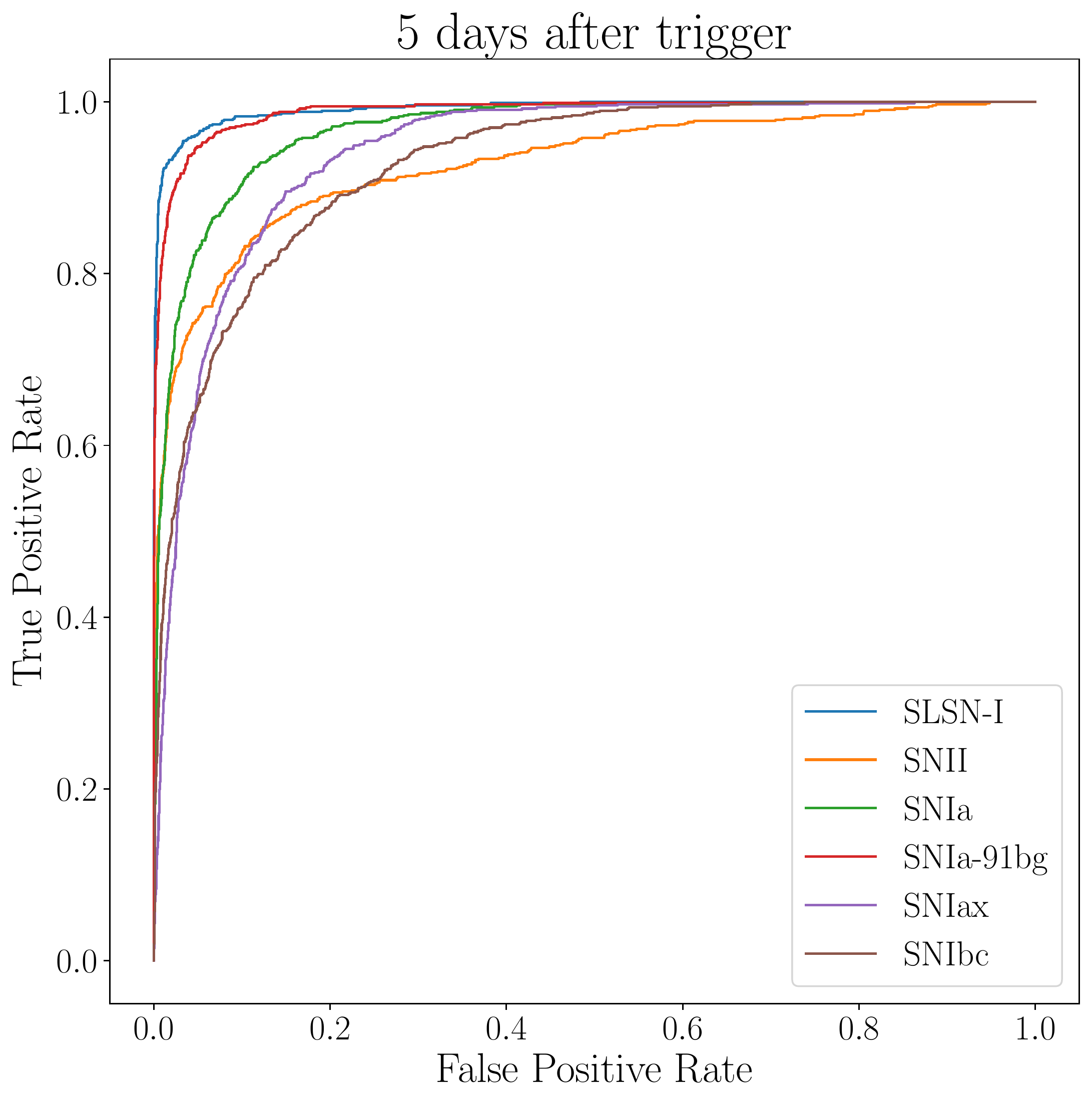}
    \includegraphics[scale=0.35,trim={0 0 0 0}]{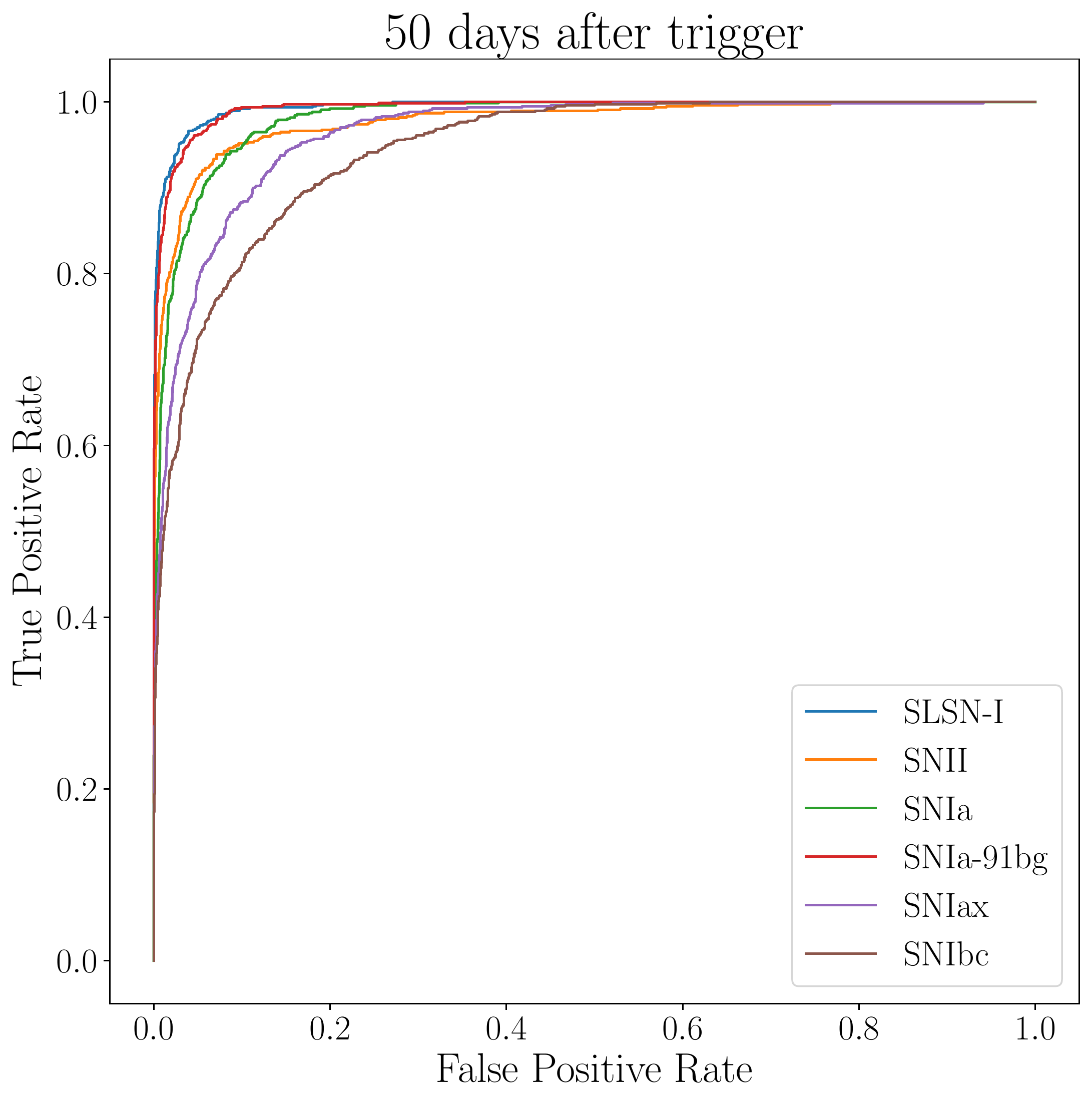}
    \includegraphics[scale=0.35,trim={-7cm 0 0 0}]{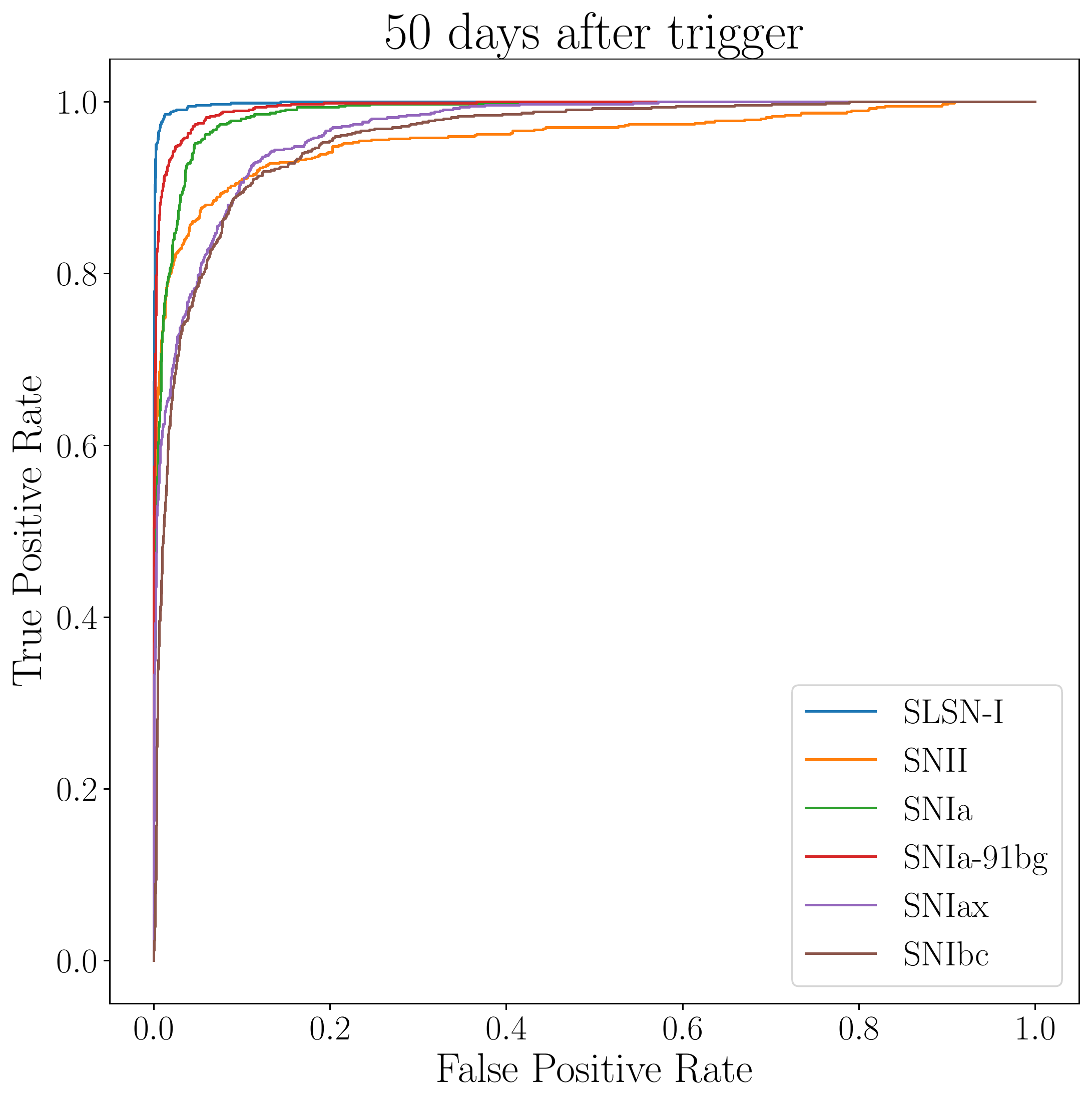}

    \centering
    \caption{Receiver operating characteristic (ROC) curves produced by \scone\ without (left) and with (right) redshift for the $t_{\mathrm{trigger}}+\{0,5,50\}$ test sets (heatmaps created from lightcurves truncated at 0, 5, and 50 days after the date of trigger).}
\end{figure*}

The datasets used for the confusion matrices in Figure~\ref{fig:cm} were also used to create ROC curves for each SN type. ROC curves for test sets without redshift are shown on the left side of Figure~\ref{fig:roc} and ROC curves for test sets with redshift are shown on the right. The addition of redshift information seems to most notably improve the model's ability to classify SLSN-I -- all three panels on the right show SLSN-I as the highest AUC curve whereas all three panels on the left show SNIa-91bg with a higher AUC curve than SLSN-I. This is consistent with our earlier observations from the confusion matrices and accuracy plots.

\begin{figure*}
    \figurenum{6}
    \label{fig:auc}
    \includegraphics[scale=0.4,trim={0 0 0 0cm}]{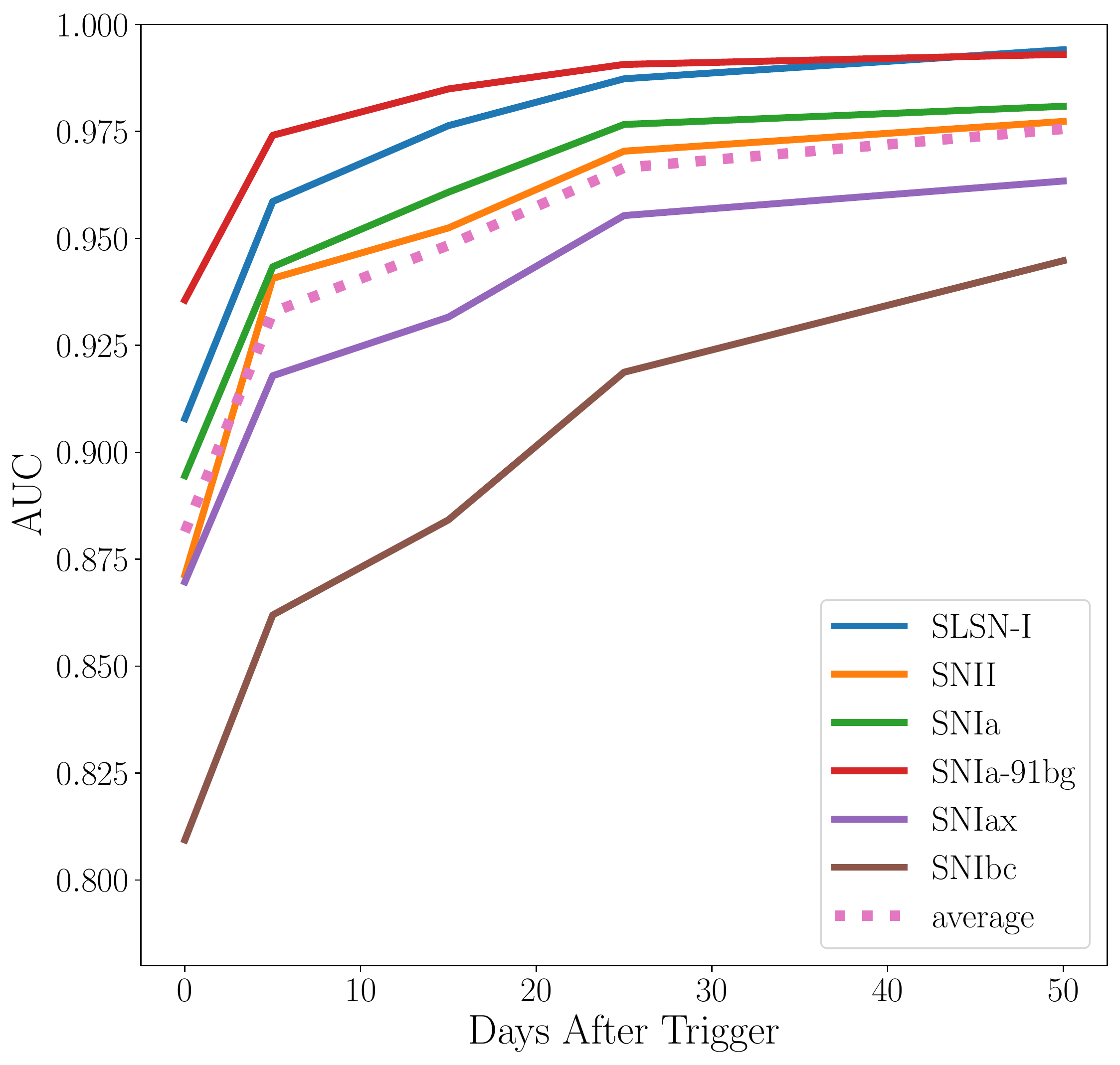}
    \includegraphics[scale=0.4,trim={0 0 0 0cm}]{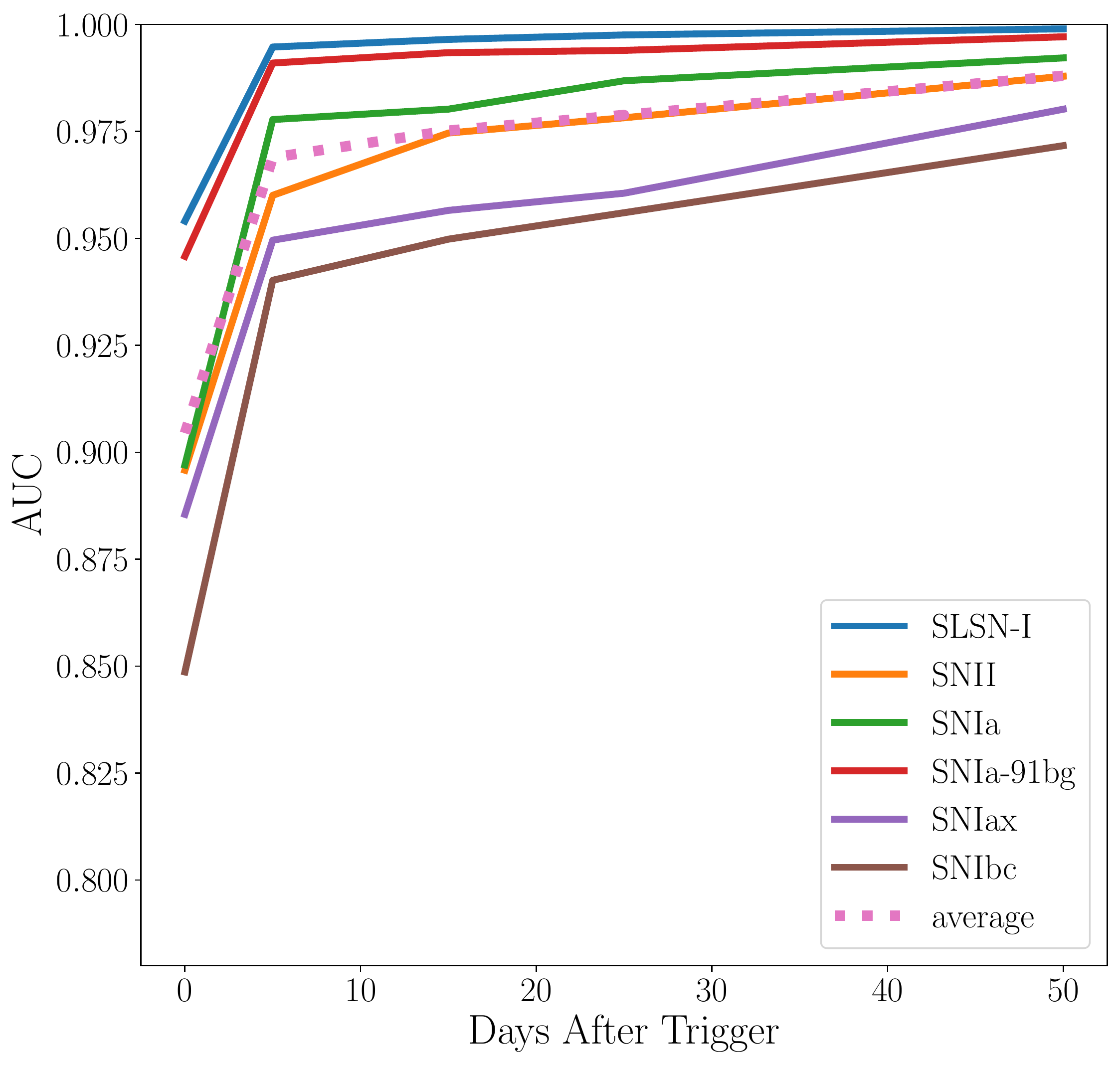}

    \centering
    \caption{Area under the ROC curve (AUC) without (left) and with (right) redshift over time for each supernova type.}
\end{figure*}

The information in the ROC curves for all $t_{\mathrm{trigger}}+N$ datasets is summarized in Figure~\ref{fig:auc}, AUC over time plots with and without redshift. The performance looks quite impressive, starting at an average AUC of above 0.9 with redshift at the date of trigger and increasing to 0.975 by 50 days after trigger. Without redshift, average AUC is still respectable, starting at 0.88 and increasing to 0.97.

\subsection{Approximating a First-Detection Trigger Definition}
Another common trigger definition used in transient surveys places the trigger at the date of the first detection ($t_{\rm first\;detection}$)rather than the second, which is the definition followed in this work. In order to more directly compare \scone's results with those of other classifiers following the first detection trigger definition, the distribution of $t_{\rm trigger}- t_{\rm first\;detection}$ was examined as well as \scone's performance on the subset of the $t_{\rm trigger}+0$ dataset with date of second detection ($t_{\rm trigger}$) at most 5 days after the date of first detection (i.e. $t_{\rm trigger}\leq t_{\rm first\;detection}+5$).

\begin{figure}
    \figurenum{7}
    \label{fig:trigger-dist}
    \centering
    \includegraphics[scale=0.6,trim={0 0 0 0cm}]{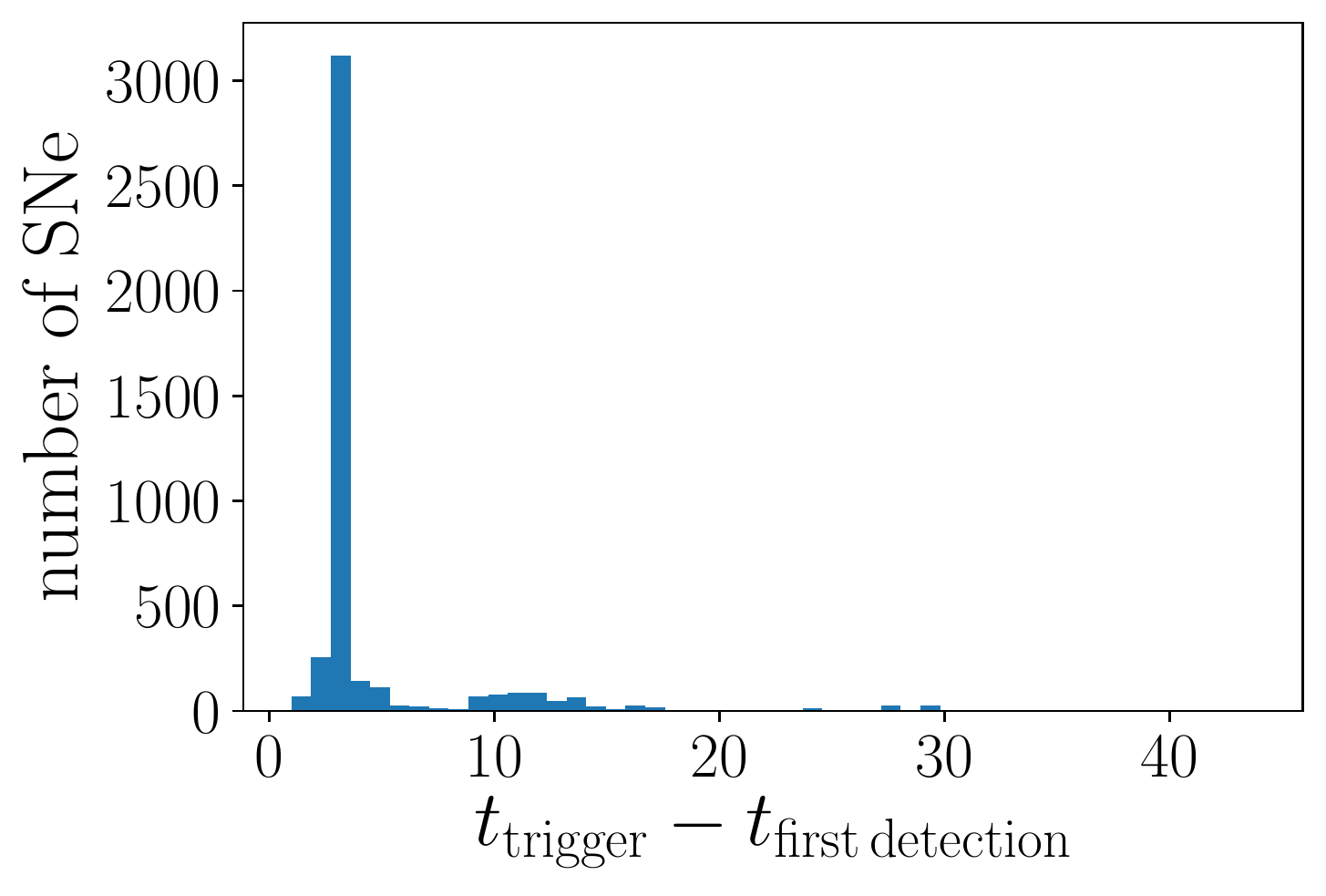}

    \caption{Distribution of $t_{\mathrm{trigger}}-t_{\mathrm{first\, detection}}$ in a \scone\ test dataset of 4608 SNe.}
\end{figure}

\begin{figure*}
    \figurenum{8}
    \label{fig:trigger}
    \includegraphics[scale=0.4,trim={0 0 0 0cm}]{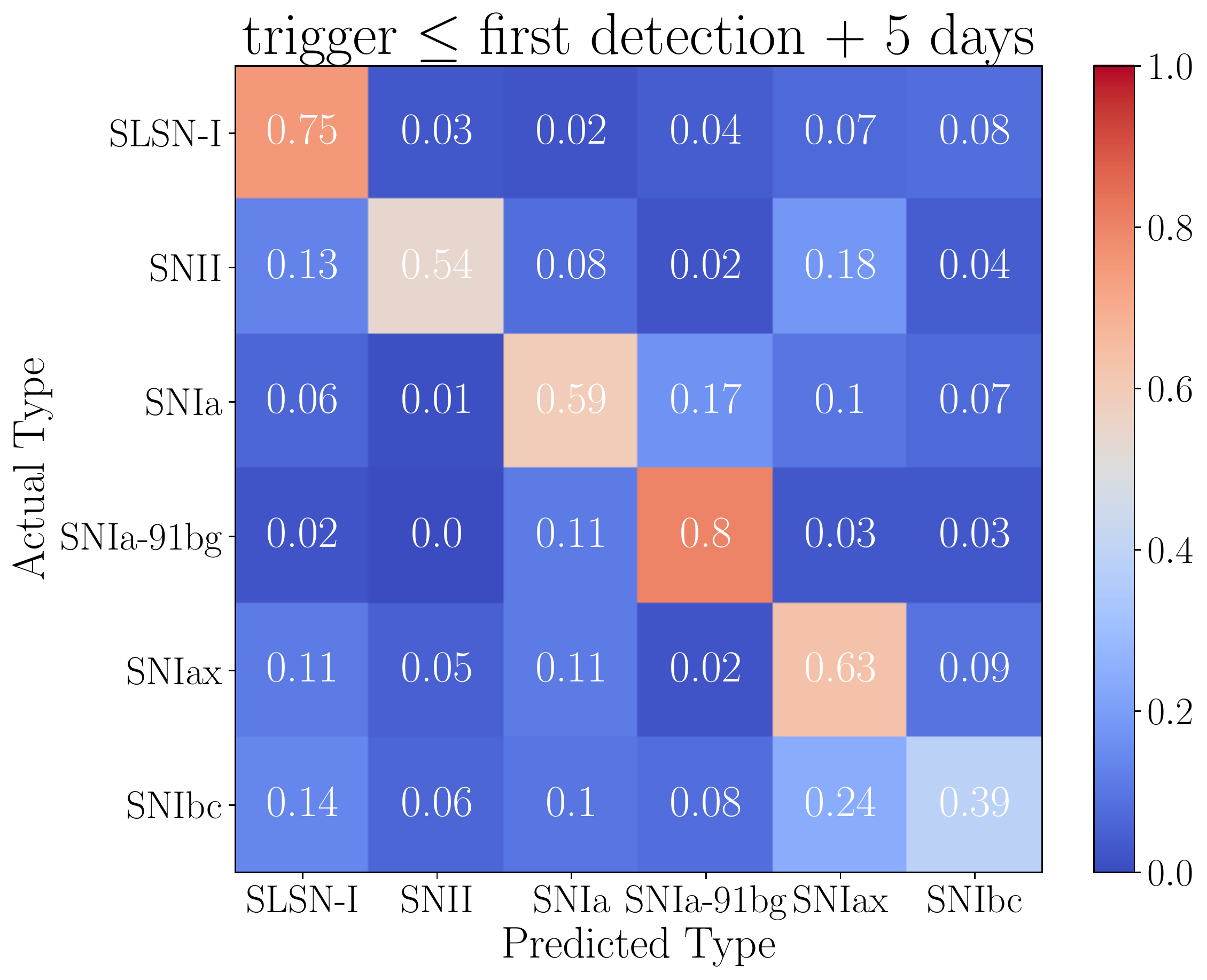}
    \includegraphics[scale=0.4,trim={0 0 0 0cm}]{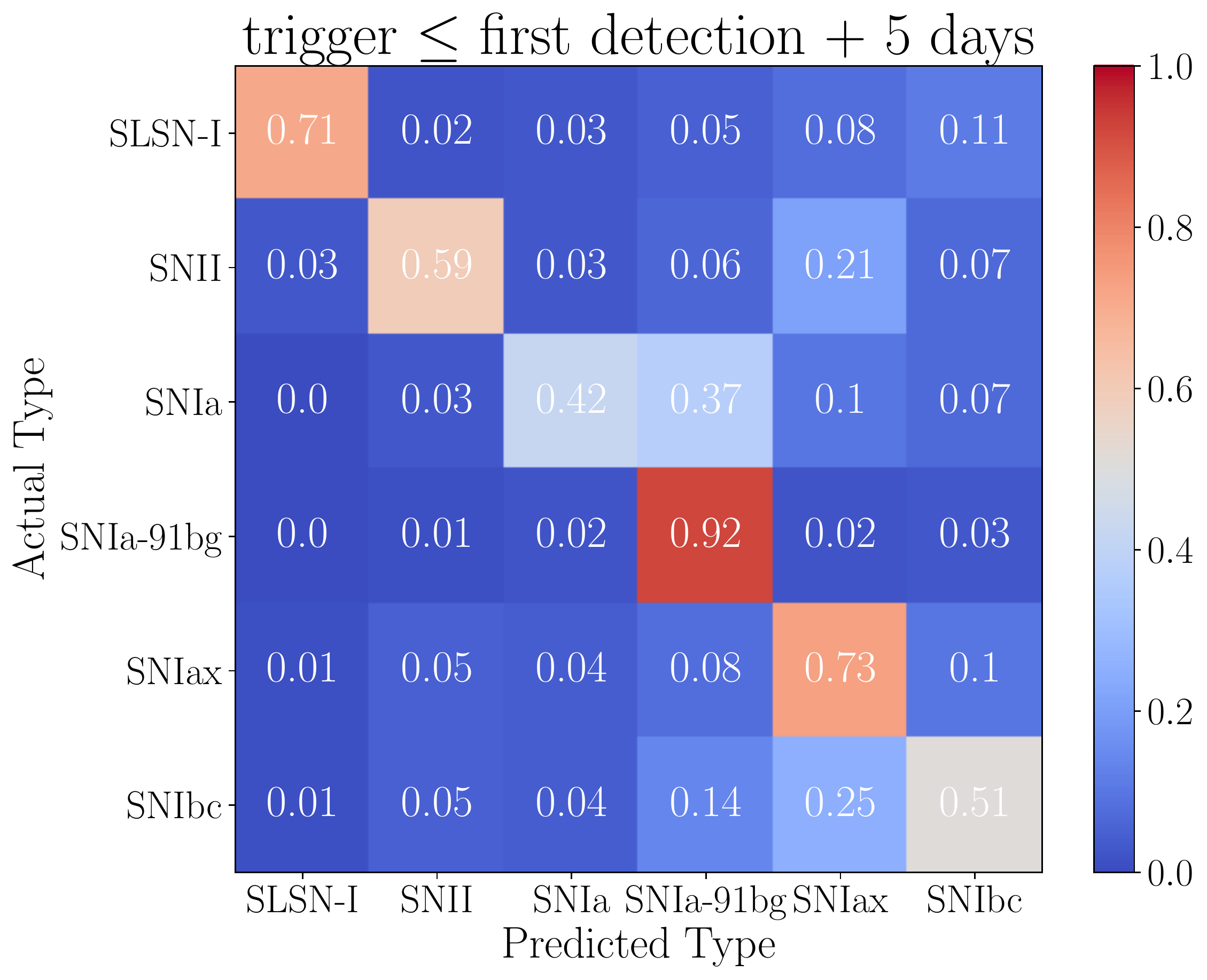}

    \caption{Normalized confusion matrices produced by \scone\ without (left) and with (right) redshift for the $t_{\rm trigger}\leq t_{\rm first\;detection}+5$ subset of the $t_{\mathrm{trigger}}+0$ test set. This cut ensures that the lightcurves used for performance evaluation are not given substantially more information than those created with the first detection trigger definition.}
\end{figure*}

Figure~\ref{fig:trigger-dist} shows that $>65$\% of $t_{\rm trigger}$ dates are no more than 5 days after the date of first detection. To further understand the direct impact of this choice of trigger definition, \scone\ was tested on the subset of the $t_{\rm trigger}+0$ dataset with date of second detection ($t_{\rm trigger}$) at most 5 days after the date of first detection. This cut ensures that the lightcurves used for classification are not given substantially more information than those created with the first detection trigger definition. The normalized confusion matrices for the $t_{\rm trigger}\leq t_{\rm first\;detection}+5$ dataset are shown with and without redshift in Figure~\ref{fig:trigger}.

With redshift, \scone's performance primarily suffers on SLSN-I and SNIa classification. SLSN-I appears to more strongly resemble SNIax and SNIbc at early times, as the SNIax confusion rose to 8\% from 1\% and the SNIbc confusion rose to 11\% from 2\%. SNIa were commonly misclassified as SNIa-91bg at early times, which is not reflected in the $t_{\rm trigger}+0$ confusion matrices in Figure~\ref{fig:cm}. Surprisingly, true SNIa-91bg were not misclassified as SNIa despite the prevalence of SNIa misclassified as SNIa-91bg. Without redshift, however, \scone's performance on the $t_{\rm trigger}\leq t_{\rm first\;detection}+5$ subset very closely resembles the $t_{\rm trigger}+0$ results shown in Figure~\ref{fig:cm}.

\subsection{Baseline Model}
A multi-layer perceptron model \citep[MLP,][]{HORNIK1989359} was developed as a baseline for direct comparison to \scone. MLP architectures are a simple type of feedforward neural network with at least 3 layers (input, hidden, output) in which each node in a particular layer is connected to every node in the subsequent layer. They have been successfully used in many general as well as image classification tasks \citep{MLPMixer,gMLP}.

The $32 \times 180 \times 2$ input heatmap is split into 180 non-overlapping ``patches" of size $32 \times 1$. The patches were chosen to be full height in the wavelength dimension to remain consistent with the full height convolutional kernels used in \scone. A $180 \times 64$-dimensional hidden layer is then computed via $h_{1,ij} = \mathrm{relu}(x^j_{i} W_{1,ji}+b_{1,j})$, where $\mathrm{relu}(x)=\mathrm{max}(0,x)$ is the rectified linear unit, $x^j$ is the $j^{\rm th}$ input heatmap patch, $W_1$ is the weight matrix learned by the network, and $b_1$ is the learned bias vector. The dimensionality of the hidden layer is then squashed to a single 64-dimensional vector with global average pooling: $h_{2,i} = \mathrm{average}(h_{1,ij})$. Finally, the output class is computed via $y_{k} = \sigma(h_{2,i} W_{2,ji}+b_{2,j})_k$, where $\sigma(\vec{x})_k=\frac{e^{x_k}}{\sum_j{e^{x_j}}}$ is the softmax function, $W_2$ is the learned weight matrix, and $b_2$ is the learned bias vector.

\begin{figure*}
    \figurenum{9}
    \label{fig:baseline}
    \includegraphics[scale=0.4,trim={0 0 0 0cm}]{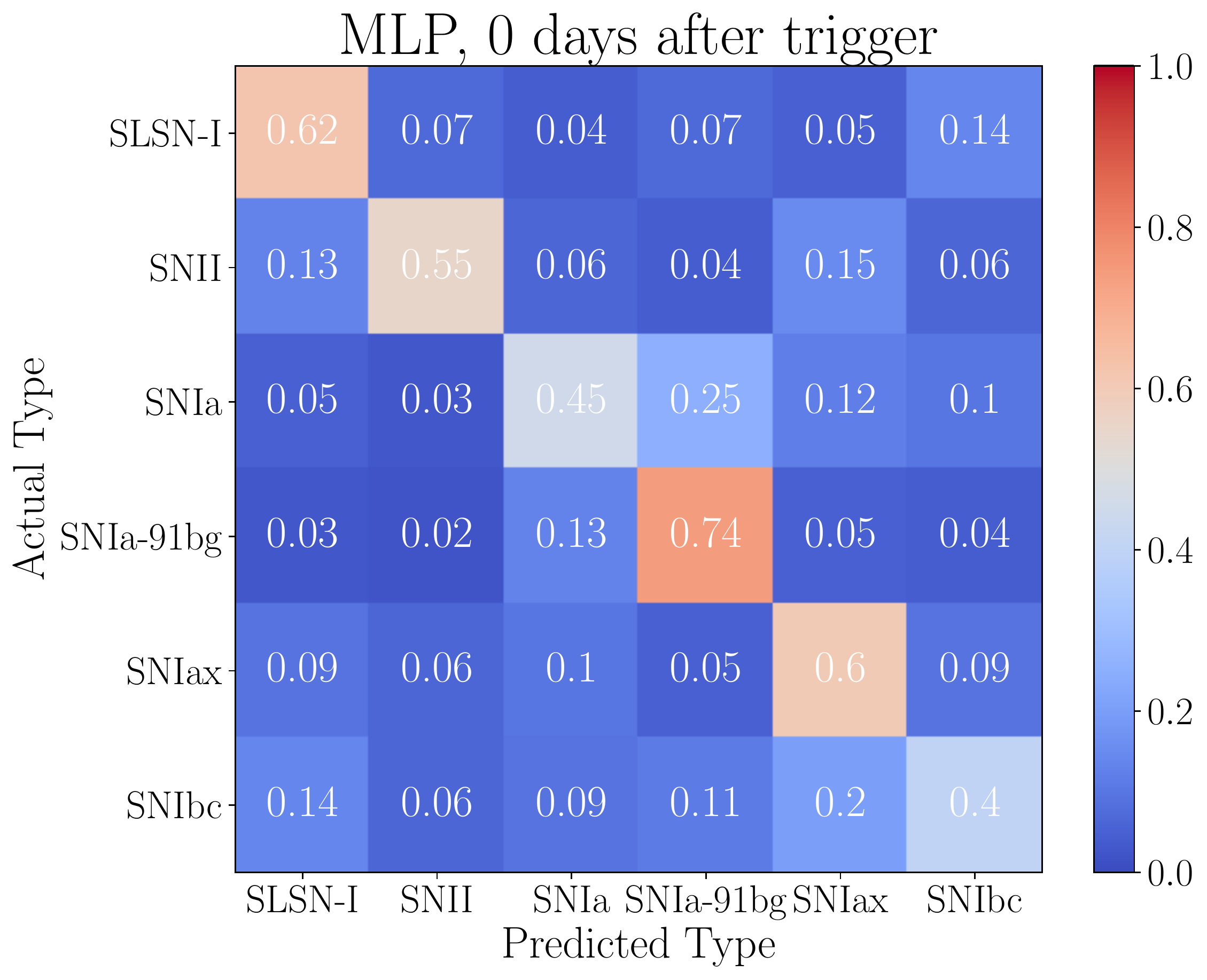}
    \includegraphics[scale=0.4,trim={0 0 0 0cm}]{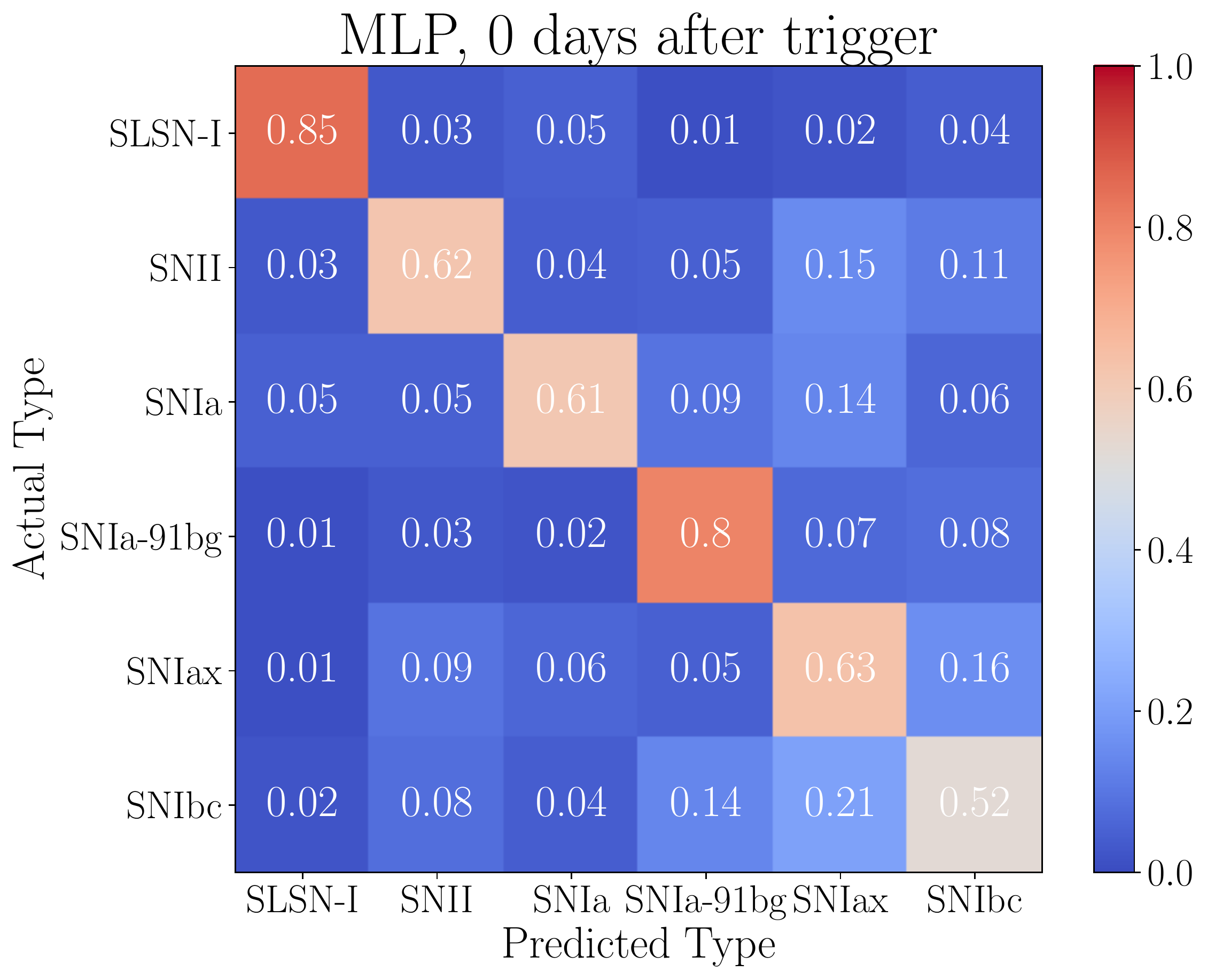}

    \centering
    \caption{Normalized confusion matrices produced by the baseline MLP model without (left) and with (right) redshift for the $t_{\mathrm{trigger}}+0$ test set (heatmaps created from lightcurves truncated at the date of trigger).}
\end{figure*}

Without redshift, our model achieved a test accuracy of 56\%. With redshift, the test accuracy improved to 67.19\%. The performance of the MLP on the $t_{\mathrm{trigger}}+0$ dataset with and without redshift is summarized in the confusion matrices in Figure~\ref{fig:baseline}. Compared to the performance of \scone\ on the $t_{\mathrm{trigger}}+0$ dataset in the top panel of Figure~\ref{fig:cm}, the MLP is less accurate at classifying most SN types, most noticeably with redshift. The degraded but still respectable performance of the MLP on classification both with and without redshift shows that these supernova types can indeed be differentiated in some hyperdimensional space by a neural network, and that \scone\ in particular possesses the required discriminatory power for this task.

\subsection{Bright Supernovae}

Bright supernovae, defined as supernovae with last included $r$-band observation $r<20$~mag, were identified from both the $t_{\mathrm{trigger}}+0$ and $t_{\mathrm{trigger}}+5$ datasets. Since fewer (and likely dimmer) observations were included for each supernova in the $t_{\mathrm{trigger}}+0$ dataset, there are much fewer examples of bright supernovae than in the $t_{\mathrm{trigger}}+5$ dataset. The bright supernovae subsets of these datasets are referred to as the ``bright $t_{\mathrm{trigger}}+N$ datasets".

\begin{figure*}
    \figurenum{10}
    \label{fig:bright}
    \includegraphics[scale=0.4,trim={0 0 0 0}]{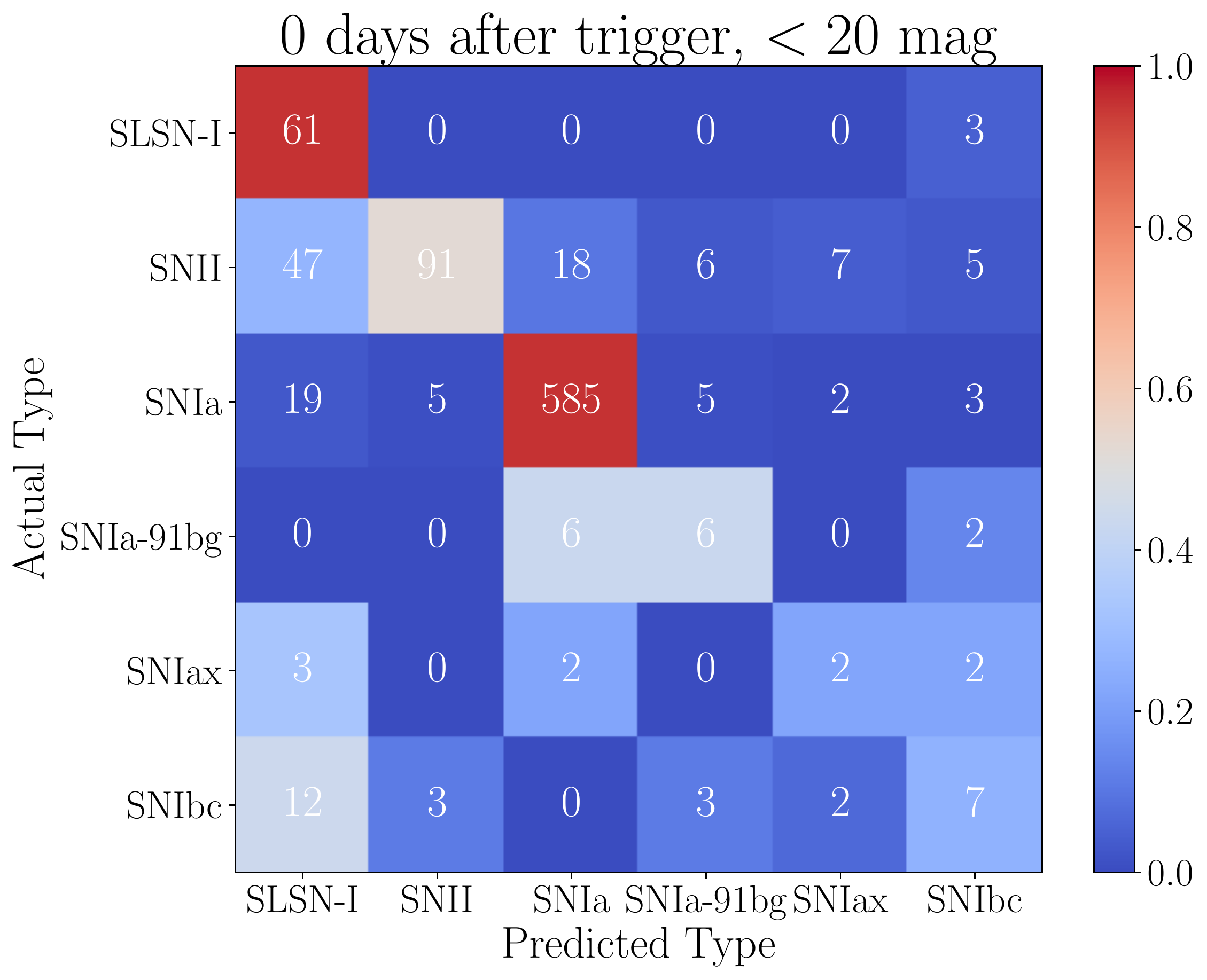}
    \includegraphics[scale=0.4,trim={0 0 0 0}]{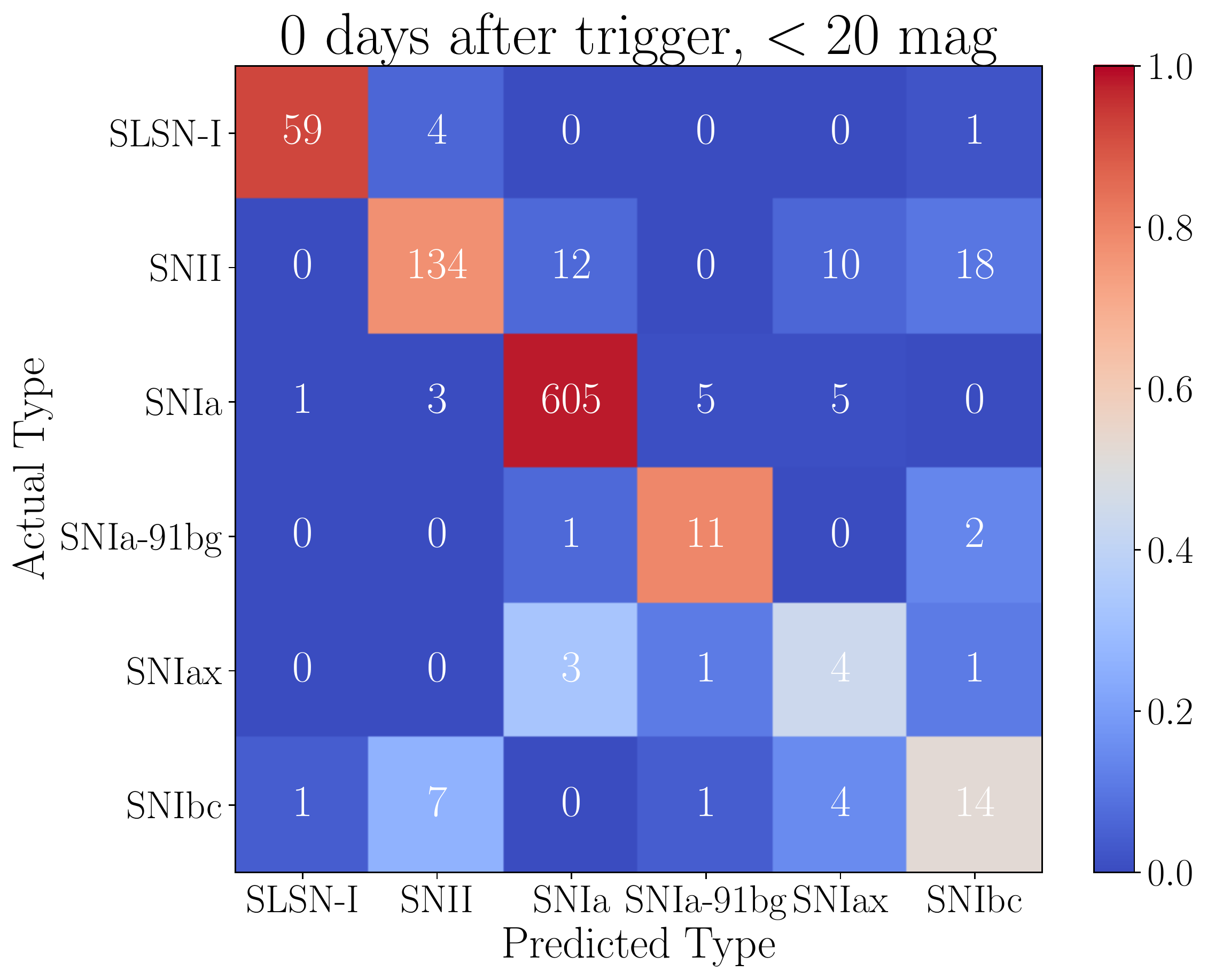}
    \includegraphics[scale=0.4,trim={0 0 0 0}]{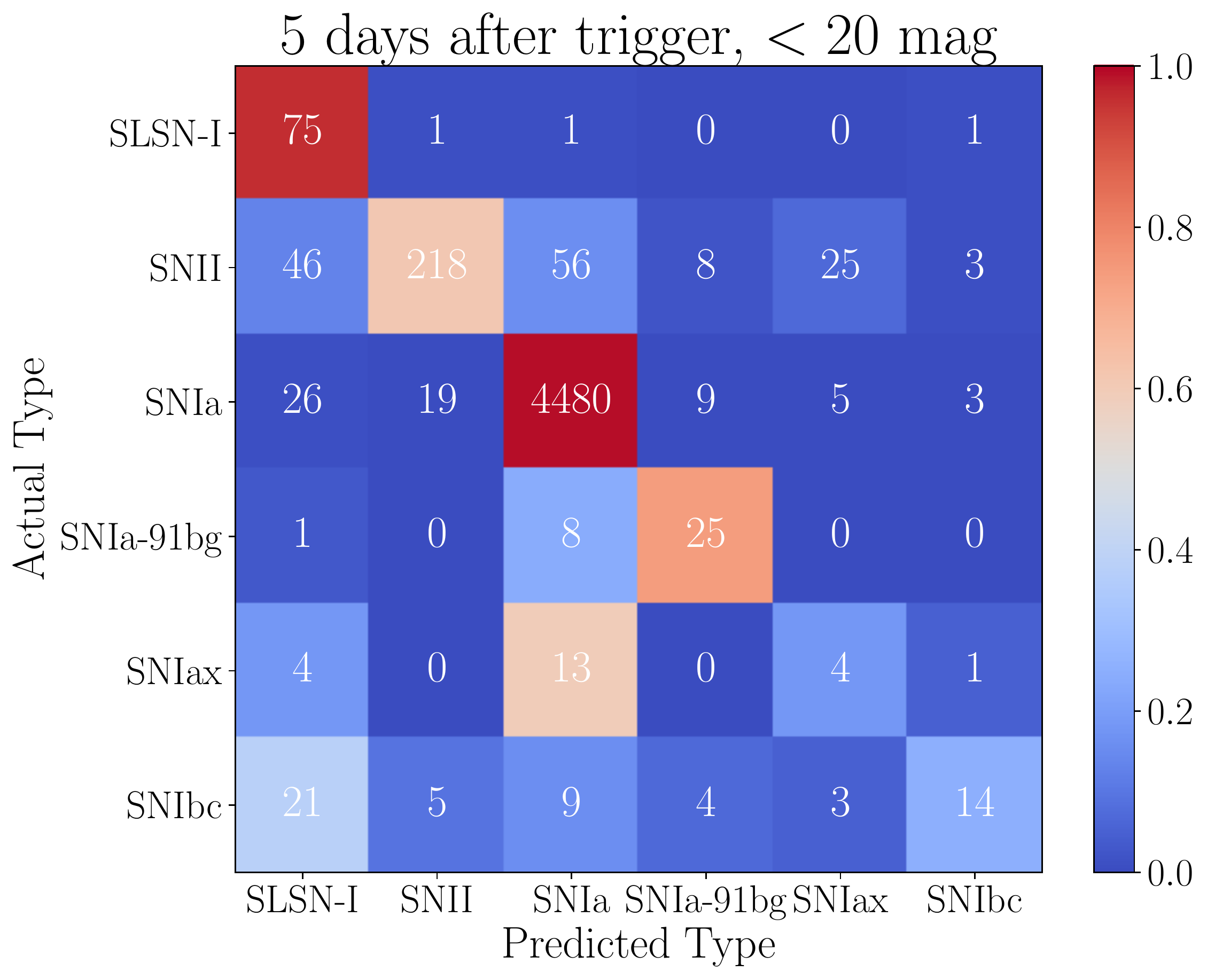}
    \includegraphics[scale=0.4,trim={0 0 0 0}]{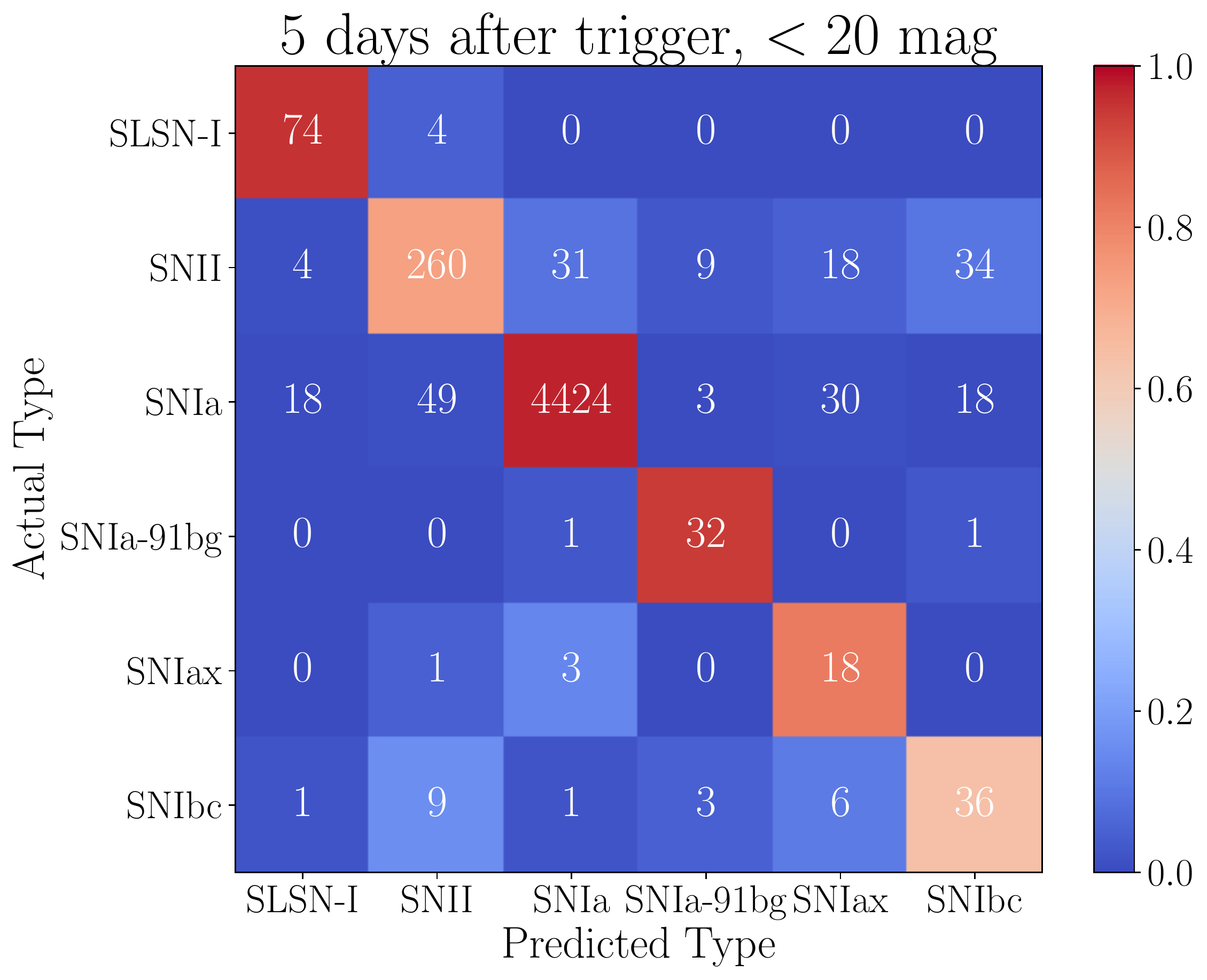}

    \centering
    \caption{Early epoch confusion matrices with (right) and without (left) redshift for the bright supernovae ($<20$ magnitude) in each $t_{\mathrm{trigger}}+N$ dataset. \scone\ was trained with a class-balanced $t_{\mathrm{trigger}}+N$ training set combined with 40\% of $<20$ magnitude supernovae. These confusion matrices were created by testing the trained \scone\ model on the full $<20$ magnitude supernovae dataset. The confusion matrices are colored according to normalized accuracies, as in Figure~\ref{fig:cm}, and are overlaid with absolute (non-normalized) values since the dataset is imbalanced.}
\end{figure*}

To evaluate the performance of \scone\ on identifying bright supernovae at early epochs, the model was trained on a regular class-balanced $t_{\mathrm{trigger}}+N$ training set, prepared as described in Section 2.4, combined with 40\% of the bright $t_{\mathrm{trigger}}+N$ dataset. The results of testing the trained \scone\ model on the bright $t_{\mathrm{trigger}}+N$ datasets are shown in Figure~\ref{fig:bright}. These confusion matrices, like the ones in Figure~\ref{fig:cm}, are colored by efficiency score. However, since the dataset is not class-balanced, the overlaid values are absolute (non-normalized) to preserve information on the relative abundance of each type. Thus, an efficiency (purity) score for each type can be calculated by dividing each main diagonal value by the sum of the values in its row (column). The overall accuracies as well as the total number of SNe in each dataset are summarized in Table~\ref{tbl:bright-acc}.

The benefits of redshift information are much more pronounced for certain types than others. As also noted in analyses of Figures~\ref{fig:cm} and ~\ref{fig:auc}, the quantity of SNIbc misclassified as SLSN-I was significantly reduced in results from \scone\ with redshift information. At the date of trigger, 44.4\% of SNIbc were misclassified as SLSN-I without redshift. This contamination rate was reduced to only 3.7\% with redshift. However, classification of bright SNIa seems relatively unaffected by the presence of redshift information. 5 days after trigger, SNIa were classified with an efficiency/accuracy of 98.6\% and a purity score of 98.1\% without redshift, and 97.4\% efficiency/accuracy and 99.1\% purity with redshift.

\begin{table}
    \centering
    \caption{Test accuracies with and without redshift information for the bright datasets.}
    \label{tbl:bright-acc}
    \begin{tabular}{l c c c}
        \hline
        & Total & Accuracy no $z$ & Accuracy with $z$\\
        \hline
        bright $t_{\mathrm{trigger}}+0$ & 907 & 82.91\% & 91.18\%\\
        bright $t_{\mathrm{trigger}}+5$ & 5088 & 94.65\% & 95.2\% \\
    \end{tabular}
\end{table}

\subsection{Mixed Dataset}

Training on the $t_{\mathrm{trigger}}+N$ datasets represents one way of deploying \scone\ for real-world transient alert applications, while training on a mixed dataset is a much less computationally expensive alternative. On one hand, testing a $t_{\mathrm{trigger}}+N$-trained model on a $t_{\mathrm{trigger}}+N$ test set yields the best classification accuracies. However, this approach requires the creation of separate datasets for each choice of $N$, which could be an expensive initial time investment depending on the number of datasets and size of each dataset (see Section 2.6 for computational requirements for heatmap creation). In this work, only five datasets ($N=0,5,15,25,50$) were created, but perhaps $N=0,1,...,50$ will be needed to accurately classify real-world transient alerts with any number of nights of photometry. Training on a mixed dataset, where each heatmap is created with a random number of nights of photometry after trigger, is a viable alternative for resource- or time-constrained applications.

\begin{figure*}
    \figurenum{11}
    \label{fig:mixed-accs}
    \includegraphics[scale=0.4,trim={0 0 0 0}]{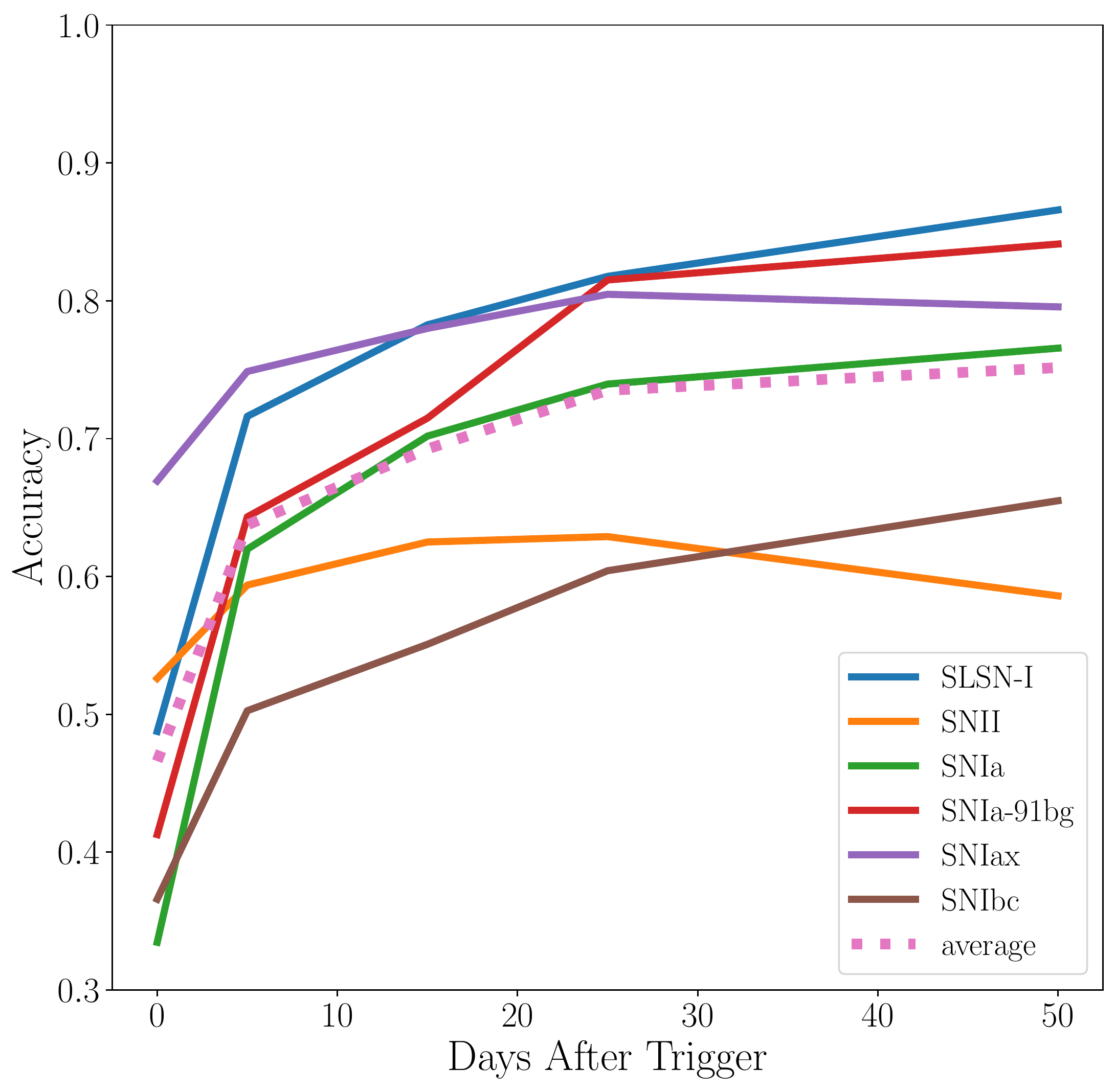}
    \includegraphics[scale=0.4,trim={0 0 0 0}]{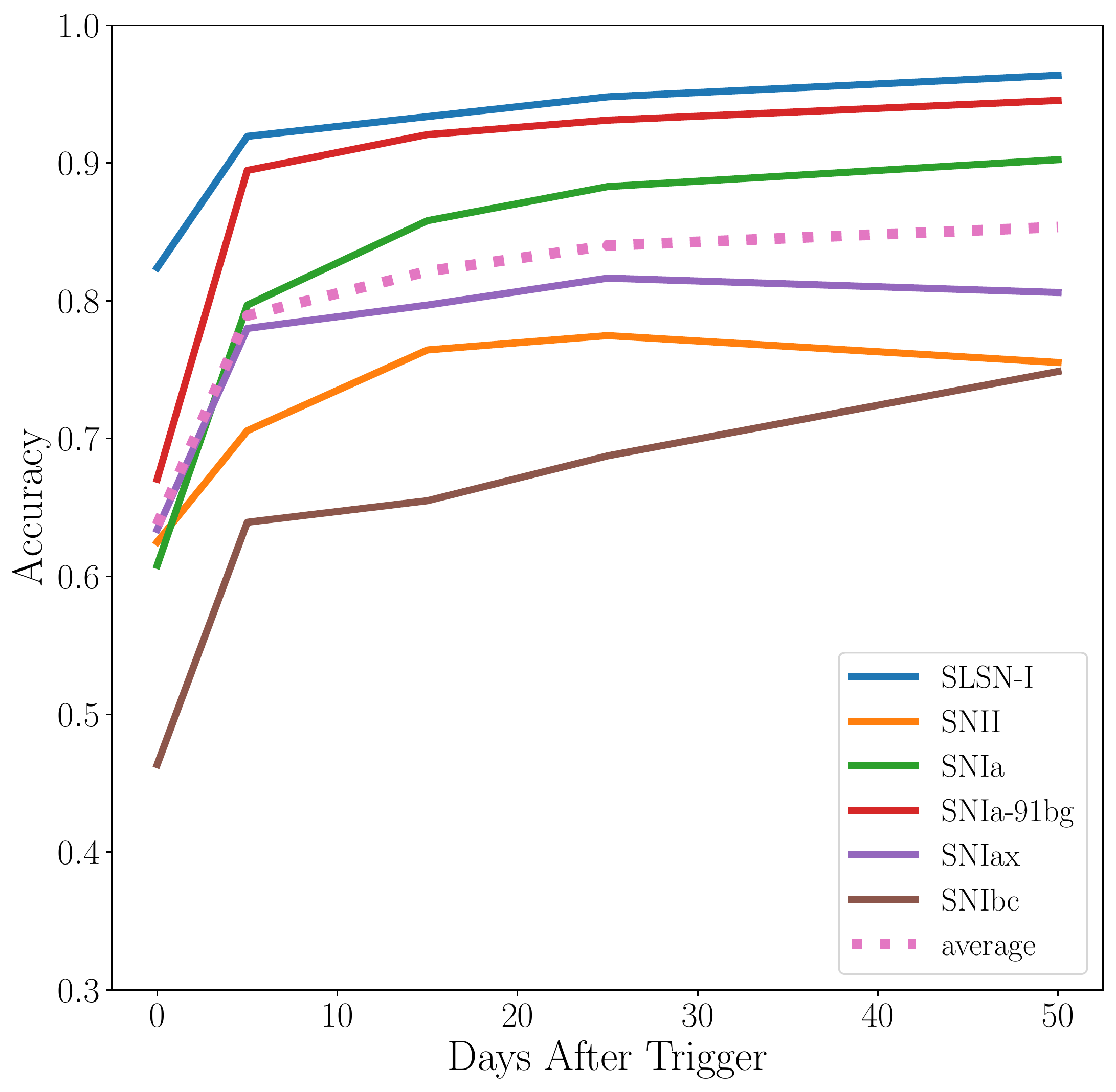}

    \centering
    \caption{Test set accuracy/efficiency without (left) and with (right) redshift over time for \scone\ trained on the mixed dataset and tested on each individual $t_{\mathrm{trigger}}+N$ dataset. The values used in these plots correspond with the diagonals on a normalized confusion matrix.}
\end{figure*}

\begin{figure*}
    \figurenum{12}
    \label{fig:mixed-auc}
    \includegraphics[scale=0.4,trim={0 0 0 0cm}]{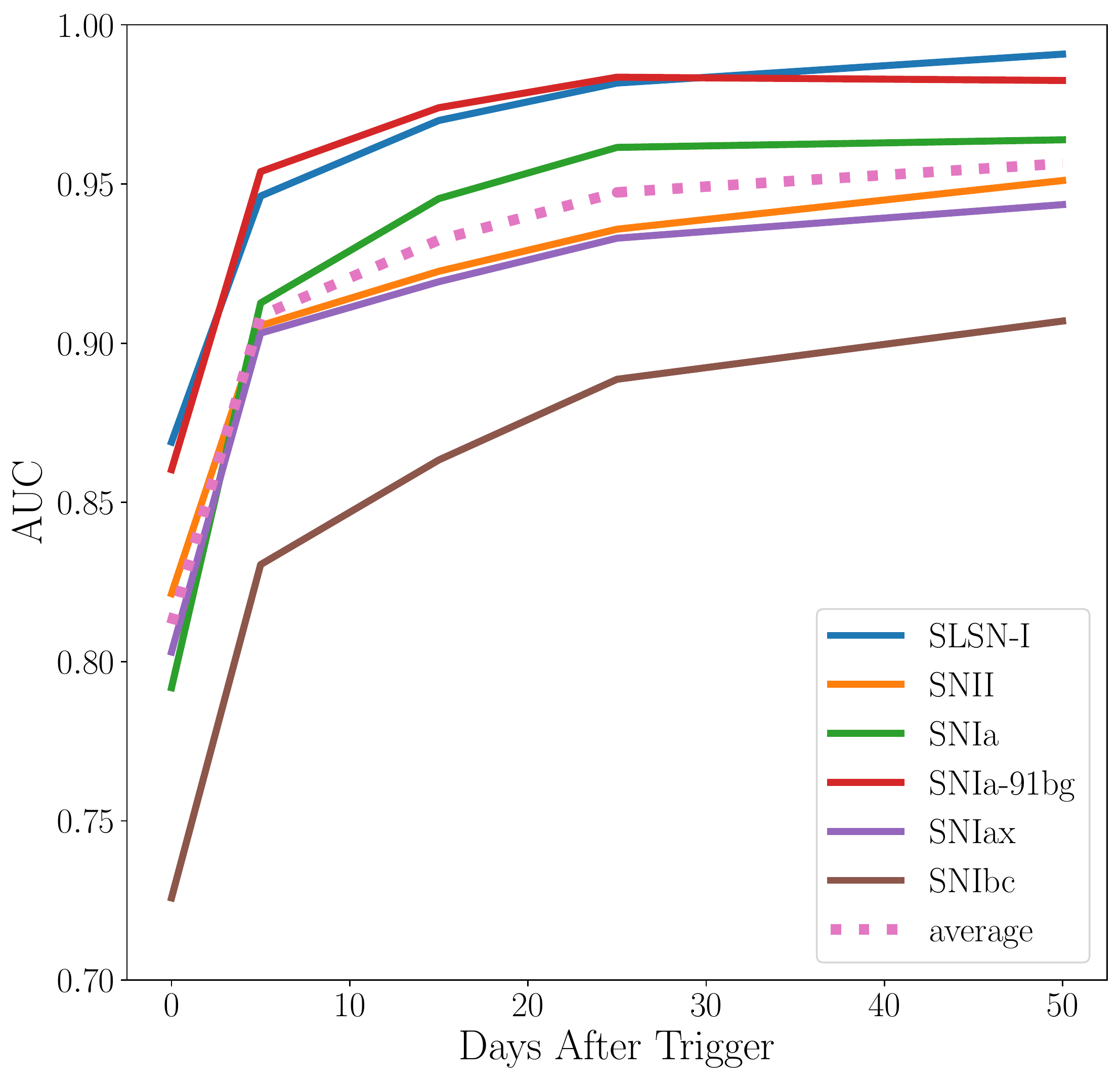}
    \includegraphics[scale=0.4,trim={0 0 0 0cm}]{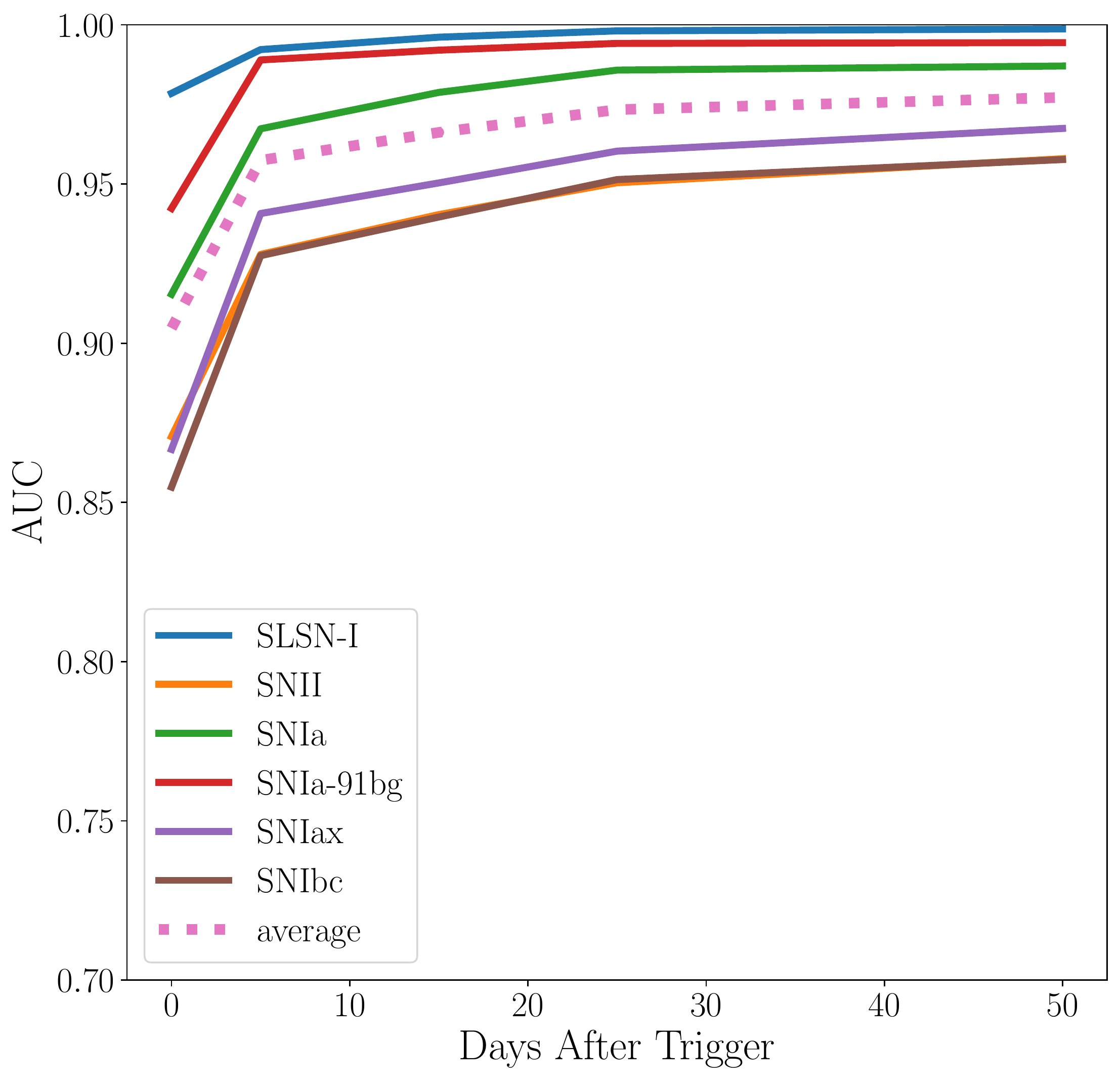}

    \centering
    \caption{Area under the ROC curve (AUC) without (left) and with (right) redshift over time for \scone\ trained on the mixed dataset and tested on each individual $t_{\mathrm{trigger}}+N$ dataset.}
\end{figure*}

To directly compare the performance of \scone\ trained on the mixed dataset and the $t_{\mathrm{trigger}}+N$ datasets, the mixed-dataset-trained model was tested on each individual $t_{\mathrm{trigger}}+N$ dataset. The accuracies over time split by SN type are summarized in Figure~\ref{fig:mixed-accs}. Compared to the results of \scone\ trained and tested on each individual $t_{\mathrm{trigger}}+N$ dataset (Figure~\ref{fig:accs}), the accuracies are lower but still respectable. The performance at the date of trigger is the most dissimilar, with average accuracy ~74\% with $z$ for a model trained on $t_{\mathrm{trigger}}+0$ and ~64\% with mixed. The performance of the mixed-trained model performs similarly to the $t_{\mathrm{trigger}}+N$-trained model by 5 days after trigger, however, both averaging just under 80\% with $z$. The AUCs over time split by SN type are shown in Figure~\ref{fig:mixed-auc}. These AUC plots are comparable to the $t_{\mathrm{trigger}}+N$ AUCs in Figure~\ref{fig:auc}, indicating that the performances of both models are comparable when averaged over all values of the prediction threshold $p$. However, the predicted class for categorical classification is not typically calculated with respect to a threshold; rather, it is defined as the class with the highest prediction confidence for each example. Thus, the AUCs are analogous to analyzing the performance on each type as its own binary classification problem, resulting in slight discrepancies from the accuracies.

\subsection{Comparison with Existing Literature}
At the time of this writing, the only work in existing literature with a similarly strong focus on early photometric classification of supernovae is RAPID \citep[][hereafter M19]{rapid}, a GRU RNN approach to photometric transient classification that differentiates between 12 transient types, including 7 supernova types. RAPID differs in several significant ways from the data, model, and results presented in this work. We highlight some of the differences in these two works below.

\subsubsection{Comparison of Methods}

The most obvious difference is in the type of neural network architecture used for classification. RAPID uses a uni-directional RNN architecture, which is designed to learn from time-series data chronologically. \scone\ employs a convolutional neural network architecture, which is most commonly used for image recognition tasks. In this instance, however, \scone\ is designed to read in data chronologically. Convolutional layers in a CNN work by computing functions on a ``sliding window" of the input image, thereby allowing the model to learn small-scale structures in the image. These windows, or the convolutional kernel, is typically a small square chosen to match the characteristic length scale of structures in the images. \scone's convolutional kernel, however, is chosen to span the full height of the input heatmap, resulting in a window that slides chronologically along the horizontal, or time, axis.

M19 trained and tested on a dataset of simulated Zwicky Transient Facility (ZTF) lightcurves, which have $g$- and $r$- band photometry, compared to the LSST lightcurves used in this work, with $ugrizY$ photometry bands. In addition to the 6 supernova types that this work focuses on, M19 includes 4 rare transient classes (pair instability supernovae (PISN), intermediate luminosity transients (ILOT), calcium-rich gap transients (CART), tidal disruption events (TDE)) as well as point-Ia and KN.

Other differences include the addition of a ``pre-explosion" class, rest- vs. observer-frame time intervals, and the choice of trigger definition. M19 chooses to include an additional class, ``pre-explosion", to describe examples at time-steps prior to the occurrence of the transient event. M19 also converts time intervals out of observer-frame by dividing by $1+z$, which is not done in this work to ensure that mistakes in redshift estimates will not be propagated to affect the lightcurve data. Finally, M19 uses the first-detection trigger date definition, while this work defines $t_{\rm trigger}$ to be the date of the second detection.

\subsubsection{Comparison of Results}
The results of \scone\ classification with redshift (right side panels of Figures~\ref{fig:cm}-\ref{fig:auc}) is used to compare with RAPID's results, as RAPID also incorporates redshift information. As described in the previous section, this work differs in many ways from M19 and the following comparison does not account for these differences; a rigorous comparison of the two models against a single dataset is left to a future work. 

Most notably, \scone\ improves upon RAPID's SNIbc and SNII classification accuracy, while RAPID performs very well at classifying early-time SNIa. From Figure 7 of M19, 12\% of SNIbc are correctly classified 2 days after detection, compared to \scone's 54\% accuracy at the date of trigger. In RAPID's results, 30\% of true SNIbc are misclassified as CART, which is not included in the datasets in this work. The second and third largest contaminants (SNIax at 19\%, then SNIa and SNIa-91bg at 8\% each), are both part of this analysis. From Figure~\ref{fig:cm}, we find that SNIax and SNIa-91bg are also major contaminants for \scone\ at 23\% and 11\%, respectively, at the date of trigger and 16\% and 4\%, respectively, 5 days after trigger. However, there is no significant contamination from SNIa, with contamination rates at 4\% on the date of trigger and 1\% 5 days after trigger.

2 days after detection, SNII is classified at 7\% accuracy by RAPID compared to 64\% accuracy at the date of trigger by \scone. The primary contaminant of SNII for RAPID 2 days after detection is SNIa at 21\%, which is not reflected in \scone's results, where the contamination rate is 6\% at the date of trigger and 3\% 5 days after trigger. The second largest contaminant, SLSN-I, is also not an issue in \scone's SNII classification. Surprisingly, the improvement over time of RAPID's SNII classification accuracy outpaces its SNIbc classification accuracy, as it is able to achieve 49\% accuracy on SNII 40 days after detection compared to 31\% accuracy on SNIbc.

While \scone's SNIa classification accuracy slowly climbs from 77\% at the date of trigger to 93\% 50 days after trigger, RAPID is able to classify SNIa at 88\% accuracy almost immediately after detection. A future direct comparison will aid in concluding whether this discrepancy is due to differences in the datasets, such as M19's exclusion of $z \geq 0.5$ objects, or something more fundamental to the model architectures.

\section{Conclusions}

Our ability to observe the universe has improved in leaps and bounds over the past century, allowing us to find new and rare transient phenomena, enrich our understanding of transient physics, and even make cosmological discoveries aided by observational data. Our photometric observing capabilities greatly outpace the rate at which we can gather the associated spectroscopic information, resulting in a vast trove of photometric data sparsely annotated by spectroscopy. In the era of large-scale sky surveys, with millions of transient alerts per night, an accurate and efficient photometric classifier is essential not only to make use of the photometric data for science analysis, but also to determine the most effective spectroscopic follow-up program early on in the life of the transient.

In this work, we presented \scone's performance classifying simulated LSST early-time supernova lightcurves for SN types Ia, II, Ibc, Ia-91bg, Iax, and SLSN-I. As a neural network-based approach, \scone\ avoids the time-intensive manual process of feature selection and engineering, and requires only raw photometric data as input. We showed that the incorporation of redshift estimates as well as errors on those estimates significantly improved classification accuracy across the board, and was especially noticeable at very early times. Notably, this is the first application of convolutional neural networks to this problem.

\scone\ was tested on 3 types of datasets: datasets of lightcurves that were truncated at 0, 5, 15, 25, and 50 days after trigger ($t_{\mathrm{trigger}}+N$ datasets); bright ($< 20$ magnitude) subsets of the $t_{\mathrm{trigger}}+\{0,5\}$ datasets; and a dataset of lightcurves truncated at a random number of nights between 0 and 50 (``mixed"). Without redshift, \scone\ was able to classify $t_{\mathrm{trigger}}+0$ lightcurves with 60\% overall accuracy, which increases to 82\% at 50 days after trigger. \scone\ with redshift information starts at 74\% overall accuracy at the date of trigger and improves to 89\% 50 days after trigger. Confusion matrices, ROC plots, and accuracy over time as well as AUC over time plots of results with and without redshift were presented to better understand classification performance and identify areas of improvement. For the bright subsets, overall accuracy is $>90$\% at the date of trigger with redshift and over 80\% without. These results improve to around 95\% accuracy both with and without redshift by 5 days after trigger. The overall accuracy over time of a mixed-dataset-trained model tested on the $t_{\mathrm{trigger}}+N$ datasets shows some degradation in accuracy at very early epochs, but may be a worthwhile lightweight alternative to the more resource-intensive process of creating many $t_{\mathrm{trigger}}+N$ datasets.

We showed that \scone's performance with redshift is competitive with existing work on early classification, such as M19, while improving on computational time requirements. \scone\ has a lightweight pre-processing step and can achieve impressive performance with a small training set. It requires only hundredths of a second to preprocess each lightcurve into a heatmap and seconds for each training epoch on GPU. This makes \scone\ a great candidate for incorporation into alert brokers for LSST and future wide-field sky surveys.

In future work, we plan to apply this model to real data to further validate the approach. We also plan to extend \scone\ to classify both full-duration and early lightcurves for more transient and variable classes in the PLAsTiCC simulations.

\section{Acknowledgments}
This work was supported by DOE grant DE-FOA-0002424, NASA Grant NNH15ZDA001N-WFIRST, and NSF grant AST-2108094. This research
used resources of the National Energy Research Scientific Computing Center (NERSC), a U.S. Department
of Energy Office of Science User Facility located at
Lawrence Berkeley National Laboratory, operated under Contract No. DE-AC02-05CH11231.





\bibliography{references}

\begin{thebibliography}{}
\expandafter\ifx\csname natexlab\endcsname\relax\def\natexlab#1{#1}\fi
\providecommand{\url}[1]{\href{#1}{#1}}
\providecommand{\dodoi}[1]{doi:~\href{http://doi.org/#1}{\nolinkurl{#1}}}
\providecommand{\doeprint}[1]{\href{http://ascl.net/#1}{\nolinkurl{http://ascl.net/#1}}}
\providecommand{\doarXiv}[1]{\href{https://arxiv.org/abs/#1}{\nolinkurl{https://arxiv.org/abs/#1}}}

\bibitem[{{Aguirre} {et~al.}(2019){Aguirre}, {Pichara}, \& {Becker}}]{aguirre}
{Aguirre}, C., {Pichara}, K., \& {Becker}, I. 2019, \mnras, 482, 5078,
  \dodoi{10.1093/mnras/sty2836}

\bibitem[{{Allam} \& {McEwen}(2021)}]{transformer}
{Allam}, Tarek, J., \& {McEwen}, J.~D. 2021, arXiv e-prints, arXiv:2105.06178.
\newblock \doarXiv{2105.06178}

\bibitem[{{Armstrong} {et~al.}(2021){Armstrong}, {Tucker}, {Rest},
  {Ridden-Harper}, {Zenati}, {Piro}, {Hinton}, {Lidman}, {Margheim}, {Narayan},
  {Shaya}, {Garnavich}, {Kasen}, {Villar}, {Zenteno}, {Arcavi}, {Drout},
  {Foley}, {Wheeler}, {Anais}, {Campillay}, {Coulter}, {Dimitriadis}, {Jones},
  {Kilpatrick}, {Mu{\~n}oz-Elgueta}, {Rojas-Bravo}, {Vargas-Gonz{\'a}lez},
  {Bulger}, {Chambers}, {Huber}, {Lowe}, {Magnier}, {Shappee}, {Smartt},
  {Smith}, {Barclay}, {Barentsen}, {Dotson}, {Gully-Santiago}, {Hedges},
  {Howell}, {Cody}, {Auchettl}, {B{\'o}di}, {Bogn{\'a}r}, {Brimacombe},
  {Brown}, {Cseh}, {Galbany}, {Hiramatsu}, {Holoien}, {Howell}, {Jha},
  {K{\"o}nyves-T{\'o}th}, {Kriskovics}, {McCully}, {Milne}, {Mu{\~n}oz}, {Pan},
  {P{\'a}l}, {Sai}, {S{\'a}rneczky}, {Smith}, {S{\'o}dor}, {Szab{\'o}},
  {Szak{\'a}ts}, {Valenti}, {Vink{\'o}}, {Wang}, {Zhang}, \& {Zsidi}}]{patrick}
{Armstrong}, P., {Tucker}, B.~E., {Rest}, A., {et~al.} 2021, \mnras, 507, 3125,
  \dodoi{10.1093/mnras/stab2138}

\bibitem[{{Boone}(2019)}]{avocado}
{Boone}, K. 2019, \aj, 158, 257, \dodoi{10.3847/1538-3881/ab5182}

\bibitem[{{Boone}(2021)}]{parsnip}
---. 2021, arXiv e-prints, arXiv:2109.13999.
\newblock \doarXiv{2109.13999}

\bibitem[{Carrasco-Davis {et~al.}(2021)Carrasco-Davis, Reyes, Valenzuela,
  Förster, Est{\'{e}}vez, Pignata, Bauer, Reyes, S{\'{a}}nchez-S{\'{a}}ez,
  Cabrera-Vives, Eyheramendy, Catelan, Arredondo, Castillo-Navarrete,
  Rodr{\'{\i}}guez-Mancini, Ruz-Mieres, Moya, Sabatini-Gacit{\'{u}}a,
  Sep{\'{u}}lveda-Cobo, Mahabal, Silva-Farf{\'{a}}n, Camacho-I{\~{n}}iguez, \&
  Galbany}]{alerce_stamp}
Carrasco-Davis, R., Reyes, E., Valenzuela, C., {et~al.} 2021, 162, 231,
  \dodoi{10.3847/1538-3881/ac0ef1}

\bibitem[{{Charnock} \& {Moss}(2017)}]{charnock_moss}
{Charnock}, T., \& {Moss}, A. 2017, \apjl, 837, L28,
  \dodoi{10.3847/2041-8213/aa603d}

\bibitem[{{Filippenko}(2005)}]{filippenko}
{Filippenko}, A.~V. 2005, in Astronomical Society of the Pacific Conference
  Series, Vol. 332, The Fate of the Most Massive Stars, ed. R.~{Humphreys} \&
  K.~{Stanek}, 34.
\newblock \doarXiv{astro-ph/0412029}

\bibitem[{{Freedman} {et~al.}(2019){Freedman}, {Madore}, {Hatt}, {Hoyt},
  {Jang}, {Beaton}, {Burns}, {Lee}, {Monson}, {Neeley}, {Phillips}, {Rich}, \&
  {Seibert}}]{freedman}
{Freedman}, W.~L., {Madore}, B.~F., {Hatt}, D., {et~al.} 2019, \apj, 882, 34,
  \dodoi{10.3847/1538-4357/ab2f73}

\bibitem[{{Frieman} {et~al.}(2008){Frieman}, {Bassett}, {Becker}, {Choi},
  {Cinabro}, {DeJongh}, {Depoy}, {Dilday}, {Doi}, {Garnavich}, {Hogan},
  {Holtzman}, {Im}, {Jha}, {Kessler}, {Konishi}, {Lampeitl}, {Marriner},
  {Marshall}, {McGinnis}, {Miknaitis}, {Nichol}, {Prieto}, {Riess}, {Richmond},
  {Romani}, {Sako}, {Schneider}, {Smith}, {Takanashi}, {Tokita}, {van der
  Heyden}, {Yasuda}, {Zheng}, {Adelman-McCarthy}, {Annis}, {Assef},
  {Barentine}, {Bender}, {Blandford}, {Boroski}, {Bremer}, {Brewington},
  {Collins}, {Crotts}, {Dembicky}, {Eastman}, {Edge}, {Edmondson}, {Elson},
  {Eyler}, {Filippenko}, {Foley}, {Frank}, {Goobar}, {Gueth}, {Gunn},
  {Harvanek}, {Hopp}, {Ihara}, {Ivezi{\'c}}, {Kahn}, {Kaplan}, {Kent},
  {Ketzeback}, {Kleinman}, {Kollatschny}, {Kron}, {Krzesi{\'n}ski}, {Lamenti},
  {Leloudas}, {Lin}, {Long}, {Lucey}, {Lupton}, {Malanushenko}, {Malanushenko},
  {McMillan}, {Mendez}, {Morgan}, {Morokuma}, {Nitta}, {Ostman}, {Pan},
  {Rockosi}, {Romer}, {Ruiz-Lapuente}, {Saurage}, {Schlesinger}, {Snedden},
  {Sollerman}, {Stoughton}, {Stritzinger}, {Subba Rao}, {Tucker}, {Vaisanen},
  {Watson}, {Watters}, {Wheeler}, {Yanny}, \& {York}}]{sdss}
{Frieman}, J.~A., {Bassett}, B., {Becker}, A., {et~al.} 2008, \aj, 135, 338,
  \dodoi{10.1088/0004-6256/135/1/338}

\bibitem[{{Guillochon} {et~al.}(2018{\natexlab{a}}){Guillochon}, {Nicholl},
  {Villar}, {Mockler}, {Narayan}, {Mandel}, {Berger}, \& {Williams}}]{SNII_3}
{Guillochon}, J., {Nicholl}, M., {Villar}, V.~A., {et~al.} 2018{\natexlab{a}},
  \\apjs, 236, 6, \dodoi{10.3847/1538-4365/aab761}

\bibitem[{{Guillochon} {et~al.}(2018{\natexlab{b}}){Guillochon}, {Nicholl},
  {Villar}, {Mockler}, {Narayan}, {Mandel}, {Berger}, \& {Williams}}]{SNIbc_3}
---. 2018{\natexlab{b}}, \\apjs, 236, 6, \dodoi{10.3847/1538-4365/aab761}

\bibitem[{{Guillochon} {et~al.}(2018{\natexlab{c}}){Guillochon}, {Nicholl},
  {Villar}, {Mockler}, {Narayan}, {Mandel}, {Berger}, \& {Williams}}]{SLSN_1}
---. 2018{\natexlab{c}}, \\apjs, 236, 6, \dodoi{10.3847/1538-4365/aab761}

\bibitem[{{Guy} {et~al.}(2010){Guy}, {Sullivan}, {Conley}, {Regnault},
  {Astier}, {Balland}, {Basa}, {Carlberg}, {Fouchez}, {Hardin}, {Hook},
  {Howell}, {Pain}, {Palanque-Delabrouille}, {Perrett}, {Pritchet}, {Rich},
  {Ruhlmann-Kleider}, {Balam}, {Baumont}, {Ellis}, {Fabbro}, {Fakhouri},
  {Fourmanoit}, {Gonz{\\\'a}lez-Gait{\\\'a}n}, {Graham}, {Hsiao}, {Kronborg},
  {Lidman}, {Mourao}, {Perlmutter}, {Ripoche}, {Suzuki}, \& {Walker}}]{SNIa_1}
{Guy}, J., {Sullivan}, M., {Conley}, A., {et~al.} 2010, \\aap, 523, A7,
  \dodoi{10.1051/0004-6361/201014468}

\bibitem[{{Hlo{\v{z}}ek} {et~al.}(2020){Hlo{\v{z}}ek}, {Ponder}, {Malz}, {Dai},
  {Narayan}, {Ishida}, {Allam}, {Bahmanyar}, {Biswas}, {Galbany}, {Jha},
  {Jones}, {Kessler}, {Lochner}, {Mahabal}, {Mandel}, {Mart{\'\i}nez-Galarza},
  {McEwen}, {Muthukrishna}, {Peiris}, {Peters}, \& {Setzer}}]{plasticc_results}
{Hlo{\v{z}}ek}, R., {Ponder}, K.~A., {Malz}, A.~I., {et~al.} 2020, arXiv
  e-prints, arXiv:2012.12392.
\newblock \doarXiv{2012.12392}

\bibitem[{{Hlozek} {et~al.}(2019){Hlozek}, {Kessler}, {Allam}, {Bahmanyar},
  {Biswas}, {Dai}, {Galbany}, {Ishida}, {Jha}, {Jones}, {Lochner}, {Mahabal},
  {Malz}, {Mandel}, {Mart{\'\i}nez-Galarza}, {McEwen}, {Muthukrishna},
  {Narayan}, {Peiris}, {Peters}, {Ponder}, {Setzer}, \& {Boucaud}}]{plasticc}
{Hlozek}, R., {Kessler}, R., {Allam}, T., {et~al.} 2019, in American
  Astronomical Society Meeting Abstracts, Vol. 233, American Astronomical
  Society Meeting Abstracts \#233, 212.01

\bibitem[{Hochreiter \& Schmidhuber(1997)}]{Hochreiter1997LongSM}
Hochreiter, S., \& Schmidhuber, J. 1997, Neural Computation, 9, 1735

\bibitem[{Hornik {et~al.}(1989)Hornik, Stinchcombe, \& White}]{HORNIK1989359}
Hornik, K., Stinchcombe, M., \& White, H. 1989, Neural Networks, 2, 359,
  \dodoi{https://doi.org/10.1016/0893-6080(89)90020-8}

\bibitem[{{Ivezi{\'c}} {et~al.}(2019){Ivezi{\'c}}, {Kahn}, {Tyson}, {Abel},
  {Acosta}, {Allsman}, {Alonso}, {AlSayyad}, {Anderson}, {Andrew}, {Angel},
  {Angeli}, {Ansari}, {Antilogus}, {Araujo}, {Armstrong}, {Arndt}, {Astier},
  {Aubourg}, {Auza}, {Axelrod}, {Bard}, {Barr}, {Barrau}, {Bartlett}, {Bauer},
  {Bauman}, {Baumont}, {Bechtol}, {Bechtol}, {Becker}, {Becla}, {Beldica},
  {Bellavia}, {Bianco}, {Biswas}, {Blanc}, {Blazek}, {Blandford}, {Bloom},
  {Bogart}, {Bond}, {Booth}, {Borgland}, {Borne}, {Bosch}, {Boutigny},
  {Brackett}, {Bradshaw}, {Brandt}, {Brown}, {Bullock}, {Burchat}, {Burke},
  {Cagnoli}, {Calabrese}, {Callahan}, {Callen}, {Carlin}, {Carlson},
  {Chandrasekharan}, {Charles-Emerson}, {Chesley}, {Cheu}, {Chiang}, {Chiang},
  {Chirino}, {Chow}, {Ciardi}, {Claver}, {Cohen-Tanugi}, {Cockrum}, {Coles},
  {Connolly}, {Cook}, {Cooray}, {Covey}, {Cribbs}, {Cui}, {Cutri}, {Daly},
  {Daniel}, {Daruich}, {Daubard}, {Daues}, {Dawson}, {Delgado}, {Dellapenna},
  {de Peyster}, {de Val-Borro}, {Digel}, {Doherty}, {Dubois},
  {Dubois-Felsmann}, {Durech}, {Economou}, {Eifler}, {Eracleous}, {Emmons},
  {Fausti Neto}, {Ferguson}, {Figueroa}, {Fisher-Levine}, {Focke}, {Foss},
  {Frank}, {Freemon}, {Gangler}, {Gawiser}, {Geary}, {Gee}, {Geha}, {Gessner},
  {Gibson}, {Gilmore}, {Glanzman}, {Glick}, {Goldina}, {Goldstein}, {Goodenow},
  {Graham}, {Gressler}, {Gris}, {Guy}, {Guyonnet}, {Haller}, {Harris},
  {Hascall}, {Haupt}, {Hernandez}, {Herrmann}, {Hileman}, {Hoblitt}, {Hodgson},
  {Hogan}, {Howard}, {Huang}, {Huffer}, {Ingraham}, {Innes}, {Jacoby}, {Jain},
  {Jammes}, {Jee}, {Jenness}, {Jernigan}, {Jevremovi{\'c}}, {Johns}, {Johnson},
  {Johnson}, {Jones}, {Juramy-Gilles}, {Juri{\'c}}, {Kalirai}, {Kallivayalil},
  {Kalmbach}, {Kantor}, {Karst}, {Kasliwal}, {Kelly}, {Kessler}, {Kinnison},
  {Kirkby}, {Knox}, {Kotov}, {Krabbendam}, {Krughoff}, {Kub{\'a}nek},
  {Kuczewski}, {Kulkarni}, {Ku}, {Kurita}, {Lage}, {Lambert}, {Lange},
  {Langton}, {Le Guillou}, {Levine}, {Liang}, {Lim}, {Lintott}, {Long},
  {Lopez}, {Lotz}, {Lupton}, {Lust}, {MacArthur}, {Mahabal}, {Mandelbaum},
  {Markiewicz}, {Marsh}, {Marshall}, {Marshall}, {May}, {McKercher}, {McQueen},
  {Meyers}, {Migliore}, {Miller}, {Mills}, {Miraval}, {Moeyens}, {Moolekamp},
  {Monet}, {Moniez}, {Monkewitz}, {Montgomery}, {Morrison}, {Mueller},
  {Muller}, {Mu{\~n}oz Arancibia}, {Neill}, {Newbry}, {Nief}, {Nomerotski},
  {Nordby}, {O'Connor}, {Oliver}, {Olivier}, {Olsen}, {O'Mullane}, {Ortiz},
  {Osier}, {Owen}, {Pain}, {Palecek}, {Parejko}, {Parsons}, {Pease},
  {Peterson}, {Peterson}, {Petravick}, {Libby Petrick}, {Petry},
  {Pierfederici}, {Pietrowicz}, {Pike}, {Pinto}, {Plante}, {Plate}, {Plutchak},
  {Price}, {Prouza}, {Radeka}, {Rajagopal}, {Rasmussen}, {Regnault}, {Reil},
  {Reiss}, {Reuter}, {Ridgway}, {Riot}, {Ritz}, {Robinson}, {Roby}, {Roodman},
  {Rosing}, {Roucelle}, {Rumore}, {Russo}, {Saha}, {Sassolas}, {Schalk},
  {Schellart}, {Schindler}, {Schmidt}, {Schneider}, {Schneider}, {Schoening},
  {Schumacher}, {Schwamb}, {Sebag}, {Selvy}, {Sembroski}, {Seppala}, {Serio},
  {Serrano}, {Shaw}, {Shipsey}, {Sick}, {Silvestri}, {Slater}, {Smith},
  {Smith}, {Sobhani}, {Soldahl}, {Storrie-Lombardi}, {Stover}, {Strauss},
  {Street}, {Stubbs}, {Sullivan}, {Sweeney}, {Swinbank}, {Szalay}, {Takacs},
  {Tether}, {Thaler}, {Thayer}, {Thomas}, {Thornton}, {Thukral}, {Tice},
  {Trilling}, {Turri}, {Van Berg}, {Vanden Berk}, {Vetter}, {Virieux},
  {Vucina}, {Wahl}, {Walkowicz}, {Walsh}, {Walter}, {Wang}, {Wang}, {Warner},
  {Wiecha}, {Willman}, {Winters}, {Wittman}, {Wolff}, {Wood-Vasey}, {Wu},
  {Xin}, {Yoachim}, \& {Zhan}}]{lsst}
{Ivezi{\'c}}, {\v{Z}}., {Kahn}, S.~M., {Tyson}, J.~A., {et~al.} 2019, \apj,
  873, 111, \dodoi{10.3847/1538-4357/ab042c}

\bibitem[{{Jha}(2017)}]{SNIax_1}
{Jha}, S.~W. 2017, {Type Iax Supernovae}, ed. A.~W. {Alsabti} \& P.~{Murdin},
  375, \dodoi{10.1007/978-3-319-21846-5_42}

\bibitem[{{Karpenka} {et~al.}(2013){Karpenka}, {Feroz}, \& {Hobson}}]{karpenka}
{Karpenka}, N.~V., {Feroz}, F., \& {Hobson}, M.~P. 2013, \mnras, 429, 1278,
  \dodoi{10.1093/mnras/sts412}

\bibitem[{{Kasen} \& {Bildsten}(2010)}]{SLSN_3}
{Kasen}, D., \& {Bildsten}, L. 2010, \\apj, 717, 245,
  \dodoi{10.1088/0004-637X/717/1/245}

\bibitem[{{Kessler} {et~al.}(2010{\natexlab{a}}){Kessler}, {Conley}, {Jha}, \&
  {Kuhlmann}}]{spcc_1}
{Kessler}, R., {Conley}, A., {Jha}, S., \& {Kuhlmann}, S. 2010{\natexlab{a}},
  arXiv e-prints, arXiv:1001.5210.
\newblock \doarXiv{1001.5210}

\bibitem[{{Kessler} {et~al.}(2009){Kessler}, {Bernstein}, {Cinabro}, {Dilday},
  {Frieman}, {Jha}, {Kuhlmann}, {Miknaitis}, {Sako}, {Taylor}, \&
  {Vanderplas}}]{snana}
{Kessler}, R., {Bernstein}, J.~P., {Cinabro}, D., {et~al.} 2009, \pasp, 121,
  1028, \dodoi{10.1086/605984}

\bibitem[{{Kessler} {et~al.}(2010{\natexlab{b}}){Kessler}, {Bassett}, {Belov},
  {Bhatnagar}, {Campbell}, {Conley}, {Frieman}, {Glazov},
  {Gonz{\'a}lez-Gait{\'a}n}, {Hlozek}, {Jha}, {Kuhlmann}, {Kunz}, {Lampeitl},
  {Mahabal}, {Newling}, {Nichol}, {Parkinson}, {Sajeeth Philip}, {Poznanski},
  {Richards}, {Rodney}, {Sako}, {Schneider}, {Smith}, {Stritzinger}, \&
  {Varughese}}]{spcc_2}
{Kessler}, R., {Bassett}, B., {Belov}, P., {et~al.} 2010{\natexlab{b}}, \pasp,
  122, 1415, \dodoi{10.1086/657607}

\bibitem[{{Kessler} {et~al.}(2010{\natexlab{c}}){Kessler}, {Bassett}, {Belov},
  {Bhatnagar}, {Campbell}, {Conley}, {Frieman}, {Glazov},
  {Gonz{\\\'a}lez-Gait{\\\'a}n}, {Hlozek}, {Jha}, {Kuhlmann}, {Kunz},
  {Lampeitl}, {Mahabal}, {Newling}, {Nichol}, {Parkinson}, {Sajeeth Philip},
  {Poznanski}, {Richards}, {Rodney}, {Sako}, {Schneider}, {Smith},
  {Stritzinger}, \& {Varughese}}]{SNII_1}
---. 2010{\natexlab{c}}, \\pasp, 122, 1415, \dodoi{10.1086/657607}

\bibitem[{{Kessler} {et~al.}(2010{\natexlab{d}}){Kessler}, {Bassett}, {Belov},
  {Bhatnagar}, {Campbell}, {Conley}, {Frieman}, {Glazov},
  {Gonz{\\\'a}lez-Gait{\\\'a}n}, {Hlozek}, {Jha}, {Kuhlmann}, {Kunz},
  {Lampeitl}, {Mahabal}, {Newling}, {Nichol}, {Parkinson}, {Sajeeth Philip},
  {Poznanski}, {Richards}, {Rodney}, {Sako}, {Schneider}, {Smith},
  {Stritzinger}, \& {Varughese}}]{SNIbc_1}
---. 2010{\natexlab{d}}, \\pasp, 122, 1415, \dodoi{10.1086/657607}

\bibitem[{{Kessler} {et~al.}(2013){Kessler}, {Guy}, {Marriner}, {Betoule},
  {Brinkmann}, {Cinabro}, {El-Hage}, {Frieman}, {Jha}, {Mosher}, \&
  {Schneider}}]{SNIa_2}
{Kessler}, R., {Guy}, J., {Marriner}, J., {et~al.} 2013, \\apj, 764, 48,
  \dodoi{10.1088/0004-637X/764/1/48}

\bibitem[{{Kessler} {et~al.}(2019){Kessler}, {Narayan}, {Avelino}, {Bachelet},
  {Biswas}, {Brown}, {Chernoff}, {Connolly}, {Dai}, {Daniel}, {Di Stefano},
  {Drout}, {Galbany}, {Gonz{\'a}lez-Gait{\'a}n}, {Graham}, {Hlo{\v{z}}ek},
  {Ishida}, {Guillochon}, {Jha}, {Jones}, {Mandel}, {Muthukrishna}, {O'Grady},
  {Peters}, {Pierel}, {Ponder}, {Pr{\v{s}}a}, {Rodney}, {Villar}, {LSST Dark
  Energy Science Collaboration}, \& {Transient and Variable Stars Science
  Collaboration}}]{plasticc-models}
{Kessler}, R., {Narayan}, G., {Avelino}, A., {et~al.} 2019, \pasp, 131, 094501,
  \dodoi{10.1088/1538-3873/ab26f1}

\bibitem[{{Kingma} \& {Ba}(2014)}]{adam}
{Kingma}, D.~P., \& {Ba}, J. 2014, arXiv e-prints, arXiv:1412.6980.
\newblock \doarXiv{1412.6980}

\bibitem[{{Kodi Ramanah} {et~al.}(2021){Kodi Ramanah}, {Arendse}, \&
  {Wojtak}}]{ramanah}
{Kodi Ramanah}, D., {Arendse}, N., \& {Wojtak}, R. 2021, arXiv e-prints,
  arXiv:2107.12399.
\newblock \doarXiv{2107.12399}

\bibitem[{Krizhevsky {et~al.}(2012)Krizhevsky, Sutskever, \& Hinton}]{alexnet}
Krizhevsky, A., Sutskever, I., \& Hinton, G.~E. 2012, Advances in neural
  information processing systems, 25, 1097

\bibitem[{LeCun {et~al.}(1989)LeCun, Boser, Denker, Henderson, Howard, Hubbard,
  \& Jackel}]{lecun1989backpropagation}
LeCun, Y., Boser, B., Denker, J.~S., {et~al.} 1989, Neural computation, 1, 541

\bibitem[{LeCun {et~al.}(1998)LeCun, Bottou, Bengio, \&
  Haffner}]{lecun1998gradient}
LeCun, Y., Bottou, L., Bengio, Y., \& Haffner, P. 1998, Proceedings of the
  IEEE, 86, 2278

\bibitem[{{Liu} {et~al.}(2021){Liu}, {Dai}, {So}, \& {Le}}]{gMLP}
{Liu}, H., {Dai}, Z., {So}, D.~R., \& {Le}, Q.~V. 2021, arXiv e-prints,
  arXiv:2105.08050.
\newblock \doarXiv{2105.08050}

\bibitem[{{Modjaz} {et~al.}(2014){Modjaz}, {Blondin}, {Kirshner}, {Matheson},
  {Berlind}, {Bianco}, {Calkins}, {Challis}, {Garnavich}, {Hicken}, {Jha},
  {Liu}, \& {Marion}}]{federica}
{Modjaz}, M., {Blondin}, S., {Kirshner}, R.~P., {et~al.} 2014, \aj, 147, 99,
  \dodoi{10.1088/0004-6256/147/5/99}

\bibitem[{{M{\"o}ller} \& {de Boissi{\`e}re}(2020)}]{moller}
{M{\"o}ller}, A., \& {de Boissi{\`e}re}, T. 2020, \mnras, 491, 4277,
  \dodoi{10.1093/mnras/stz3312}

\bibitem[{{Moss}(2018)}]{moss}
{Moss}, A. 2018, arXiv e-prints, arXiv:1810.06441.
\newblock \doarXiv{1810.06441}

\bibitem[{{Muthukrishna} {et~al.}(2019){Muthukrishna}, {Narayan}, {Mandel},
  {Biswas}, \& {Hlo{\v{z}}ek}}]{rapid}
{Muthukrishna}, D., {Narayan}, G., {Mandel}, K.~S., {Biswas}, R., \&
  {Hlo{\v{z}}ek}, R. 2019, \pasp, 131, 118002, \dodoi{10.1088/1538-3873/ab1609}

\bibitem[{{Naul} {et~al.}(2018){Naul}, {Bloom}, {P{\'e}rez}, \& {van der
  Walt}}]{naul}
{Naul}, B., {Bloom}, J.~S., {P{\'e}rez}, F., \& {van der Walt}, S. 2018, Nature
  Astronomy, 2, 151, \dodoi{10.1038/s41550-017-0321-z}

\bibitem[{{Nicholl} {et~al.}(2017){Nicholl}, {Guillochon}, \&
  {Berger}}]{SLSN_2}
{Nicholl}, M., {Guillochon}, J., \& {Berger}, E. 2017, \\apj, 850, 55,
  \dodoi{10.3847/1538-4357/aa9334}

\bibitem[{{Pasquet} {et~al.}(2019){Pasquet}, {Pasquet}, {Chaumont}, \&
  {Fouchez}}]{pelican}
{Pasquet}, J., {Pasquet}, J., {Chaumont}, M., \& {Fouchez}, D. 2019, \aap, 627,
  A21, \dodoi{10.1051/0004-6361/201834473}

\bibitem[{{Perets} {et~al.}(2010){Perets}, {Gal-Yam}, {Mazzali}, {Arnett},
  {Kagan}, {Filippenko}, {Li}, {Arcavi}, {Cenko}, {Fox}, {Leonard}, {Moon},
  {Sand}, {Soderberg}, {Anderson}, {James}, {Foley}, {Ganeshalingam}, {Ofek},
  {Bildsten}, {Nelemans}, {Shen}, {Weinberg}, {Metzger}, {Piro}, {Quataert},
  {Kiewe}, \& {Poznanski}}]{perets}
{Perets}, H.~B., {Gal-Yam}, A., {Mazzali}, P.~A., {et~al.} 2010, \nat, 465,
  322, \dodoi{10.1038/nature09056}

\bibitem[{{Perlmutter} {et~al.}(1999){Perlmutter}, {Turner}, \&
  {White}}]{perlmutter}
{Perlmutter}, S., {Turner}, M.~S., \& {White}, M. 1999, \prl, 83, 670,
  \dodoi{10.1103/PhysRevLett.83.670}

\bibitem[{{Pierel} {et~al.}(2018{\natexlab{a}}){Pierel}, {Rodney}, {Avelino},
  {Bianco}, {Filippenko}, {Foley}, {Friedman}, {Hicken}, {Hounsell}, {Jha},
  {Kessler}, {Kirshner}, {Mandel}, {Narayan}, {Scolnic}, \&
  {Strolger}}]{SNIa_3}
{Pierel}, J.~D.~R., {Rodney}, S., {Avelino}, A., {et~al.} 2018{\natexlab{a}},
  \\pasp, 130, 114504, \dodoi{10.1088/1538-3873/aadb7a}

\bibitem[{{Pierel} {et~al.}(2018{\natexlab{b}}){Pierel}, {Rodney}, {Avelino},
  {Bianco}, {Filippenko}, {Foley}, {Friedman}, {Hicken}, {Hounsell}, {Jha},
  {Kessler}, {Kirshner}, {Mandel}, {Narayan}, {Scolnic}, \&
  {Strolger}}]{SNII_2}
---. 2018{\natexlab{b}}, \\pasp, 130, 114504, \dodoi{10.1088/1538-3873/aadb7a}

\bibitem[{{Pierel} {et~al.}(2018{\natexlab{c}}){Pierel}, {Rodney}, {Avelino},
  {Bianco}, {Filippenko}, {Foley}, {Friedman}, {Hicken}, {Hounsell}, {Jha},
  {Kessler}, {Kirshner}, {Mandel}, {Narayan}, {Scolnic}, \&
  {Strolger}}]{SNIbc_2}
---. 2018{\natexlab{c}}, \\pasp, 130, 114504, \dodoi{10.1088/1538-3873/aadb7a}

\bibitem[{{Poznanski} {et~al.}(2007){Poznanski}, {Maoz}, \&
  {Gal-Yam}}]{poznanski}
{Poznanski}, D., {Maoz}, D., \& {Gal-Yam}, A. 2007, \aj, 134, 1285,
  \dodoi{10.1086/520956}

\bibitem[{{Pursiainen} {et~al.}(2018){Pursiainen}, {Childress}, {Smith},
  {Prajs}, {Sullivan}, {Davis}, {Foley}, {Asorey}, {Calcino}, {Carollo},
  {Curtin}, {D'Andrea}, {Glazebrook}, {Gutierrez}, {Hinton}, {Hoormann},
  {Inserra}, {Kessler}, {King}, {Kuehn}, {Lewis}, {Lidman}, {Macaulay},
  {M{\"o}ller}, {Nichol}, {Sako}, {Sommer}, {Swann}, {Tucker}, {Uddin},
  {Wiseman}, {Zhang}, {Abbott}, {Abdalla}, {Allam}, {Annis}, {Avila}, {Brooks},
  {Buckley-Geer}, {Burke}, {Carnero Rosell}, {Carrasco Kind}, {Carretero},
  {Castander}, {Cunha}, {Davis}, {De Vicente}, {Diehl}, {Doel}, {Eifler},
  {Flaugher}, {Fosalba}, {Frieman}, {Garc{\'\i}a-Bellido}, {Gruen}, {Gruendl},
  {Gutierrez}, {Hartley}, {Hollowood}, {Honscheid}, {James}, {Jeltema},
  {Kuropatkin}, {Li}, {Lima}, {Maia}, {Martini}, {Menanteau}, {Ogando},
  {Plazas}, {Roodman}, {Sanchez}, {Scarpine}, {Schindler}, {Smith},
  {Soares-Santos}, {Sobreira}, {Suchyta}, {Swanson}, {Tarle}, {Tucker},
  {Walker}, \& {DES Collaboration}}]{pursiainen}
{Pursiainen}, M., {Childress}, M., {Smith}, M., {et~al.} 2018, \mnras, 481,
  894, \dodoi{10.1093/mnras/sty2309}

\bibitem[{Qu(2021)}]{helen_qu_2021_5602043}
Qu, H. 2021, {helenqu/scone: early lightcurve classification release}, v1.1.0,
  Zenodo, \dodoi{10.5281/zenodo.5602043}

\bibitem[{Qu {et~al.}(2021)Qu, Sako, Möller, \& Doux}]{Qu_2021}
Qu, H., Sako, M., Möller, A., \& Doux, C. 2021, The Astronomical Journal, 162,
  67, \dodoi{10.3847/1538-3881/ac0824}

\bibitem[{{Richards} {et~al.}(2012){Richards}, {Homrighausen}, {Freeman},
  {Schafer}, \& {Poznanski}}]{richards}
{Richards}, J.~W., {Homrighausen}, D., {Freeman}, P.~E., {Schafer}, C.~M., \&
  {Poznanski}, D. 2012, \mnras, 419, 1121,
  \dodoi{10.1111/j.1365-2966.2011.19768.x}

\bibitem[{{Riess}(1998)}]{riess}
{Riess}, A.~G. 1998, in American Astronomical Society Meeting Abstracts, Vol.
  192, American Astronomical Society Meeting Abstracts \#192, 17.06

\bibitem[{{Riess} {et~al.}(2019){Riess}, {Casertano}, {Yuan}, {Macri}, \&
  {Scolnic}}]{riess_2019}
{Riess}, A.~G., {Casertano}, S., {Yuan}, W., {Macri}, L.~M., \& {Scolnic}, D.
  2019, \apj, 876, 85, \dodoi{10.3847/1538-4357/ab1422}

\bibitem[{{Sako} {et~al.}(2008){Sako}, {Bassett}, {Becker}, {Cinabro},
  {DeJongh}, {Depoy}, {Dilday}, {Doi}, {Frieman}, {Garnavich}, {Hogan},
  {Holtzman}, {Jha}, {Kessler}, {Konishi}, {Lampeitl}, {Marriner}, {Miknaitis},
  {Nichol}, {Prieto}, {Riess}, {Richmond}, {Romani}, {Schneider}, {Smith},
  {SubbaRao}, {Takanashi}, {Tokita}, {van der Heyden}, {Yasuda}, {Zheng},
  {Barentine}, {Brewington}, {Choi}, {Dembicky}, {Harnavek}, {Ihara}, {Im},
  {Ketzeback}, {Kleinman}, {Krzesi{\'n}ski}, {Long}, {Malanushenko},
  {Malanushenko}, {McMillan}, {Morokuma}, {Nitta}, {Pan}, {Saurage}, \&
  {Snedden}}]{sako2008}
{Sako}, M., {Bassett}, B., {Becker}, A., {et~al.} 2008, \aj, 135, 348,
  \dodoi{10.1088/0004-6256/135/1/348}

\bibitem[{{Sako} {et~al.}(2011){Sako}, {Bassett}, {Connolly}, {Dilday},
  {Cambell}, {Frieman}, {Gladney}, {Kessler}, {Lampeitl}, {Marriner}, {Miquel},
  {Nichol}, {Schneider}, {Smith}, \& {Sollerman}}]{sako2011}
{Sako}, M., {Bassett}, B., {Connolly}, B., {et~al.} 2011, \apj, 738, 162,
  \dodoi{10.1088/0004-637X/738/2/162}

\bibitem[{{S{\'a}nchez-S{\'a}ez} {et~al.}(2021){S{\'a}nchez-S{\'a}ez}, {Reyes},
  {Valenzuela}, {F{\"o}rster}, {Eyheramendy}, {Elorrieta}, {Bauer},
  {Cabrera-Vives}, {Est{\'e}vez}, {Catelan}, {Pignata}, {Huijse}, {De Cicco},
  {Ar{\'e}valo}, {Carrasco-Davis}, {Abril}, {Kurtev}, {Borissova}, {Arredondo},
  {Castillo-Navarrete}, {Rodriguez}, {Ruz-Mieres}, {Moya},
  {Sabatini-Gacit{\'u}a}, {Sep{\'u}lveda-Cobo}, \&
  {Camacho-I{\~n}iguez}}]{alerce_lc}
{S{\'a}nchez-S{\'a}ez}, P., {Reyes}, I., {Valenzuela}, C., {et~al.} 2021, \aj,
  161, 141, \dodoi{10.3847/1538-3881/abd5c1}

\bibitem[{{Smith} {et~al.}(2020){Smith}, {Sullivan}, {Wiseman}, {Kessler},
  {Scolnic}, {Brout}, {D'Andrea}, {Davis}, {Foley}, {Frohmaier}, {Galbany},
  {Gupta}, {Guti{\'e}rrez}, {Hinton}, {Kelsey}, {Lidman}, {Macaulay},
  {M{\"o}ller}, {Nichol}, {Nugent}, {Palmese}, {Pursiainen}, {Sako}, {Swann},
  {Thomas}, {Tucker}, {Vincenzi}, {Carollo}, {Lewis}, {Sommer}, {Abbott},
  {Aguena}, {Allam}, {Avila}, {Bertin}, {Bhargava}, {Brooks}, {Buckley-Geer},
  {Burke}, {Carnero Rosell}, {Carrasco Kind}, {Costanzi}, {da Costa}, {De
  Vicente}, {Desai}, {Diehl}, {Doel}, {Eifler}, {Everett}, {Flaugher},
  {Fosalba}, {Frieman}, {Garc{\'\i}a-Bellido}, {Gaztanaga}, {Glazebrook},
  {Gruen}, {Gruendl}, {Gschwend}, {Gutierrez}, {Hartley}, {Hollowood},
  {Honscheid}, {James}, {Krause}, {Kuehn}, {Kuropatkin}, {Lima}, {MacCrann},
  {Maia}, {Marshall}, {Martini}, {Melchior}, {Menanteau}, {Miquel},
  {Paz-Chinch{\'o}n}, {Plazas}, {Romer}, {Roodman}, {Rykoff}, {Sanchez},
  {Scarpine}, {Schubnell}, {Serrano}, {Sevilla-Noarbe}, {Suchyta}, {Swanson},
  {Tarle}, {Thomas}, {Tucker}, {Varga}, {Walker}, \& {DES Collaboration}}]{des}
{Smith}, M., {Sullivan}, M., {Wiseman}, P., {et~al.} 2020, \mnras, 494, 4426,
  \dodoi{10.1093/mnras/staa946}

\bibitem[{{Sollerman} {et~al.}(2021){Sollerman}, {Yang}, {Schulze},
  {Strotjohann}, {Jerkstrand}, {Van Dyk}, {Kool}, {Barbarino}, {Brink},
  {Bruch}, {De}, {Filippenko}, {Fremling}, {Patra}, {Perley}, {Yan}, {Yang},
  {Andreoni}, {Campbell}, {Coughlin}, {Kasliwal}, {Kim}, {Rigault}, {Shin},
  {Tzanidakis}, {Ashley}, {Moore}, \& {Travouillon}}]{sollerman}
{Sollerman}, J., {Yang}, S., {Schulze}, S., {et~al.} 2021, arXiv e-prints,
  arXiv:2107.14503.
\newblock \doarXiv{2107.14503}

\bibitem[{{Sullivan} {et~al.}(2006){Sullivan}, {Howell}, {Perrett}, {Nugent},
  {Astier}, {Aubourg}, {Balam}, {Basa}, {Carlberg}, {Conley}, {Fabbro},
  {Fouchez}, {Guy}, {Hook}, {Lafoux}, {Neill}, {Pain}, {Palanque-Delabrouille},
  {Pritchet}, {Regnault}, {Rich}, {Taillet}, {Aldering}, {Baumont}, {Bronder},
  {Filiol}, {Knop}, {Perlmutter}, \& {Tao}}]{sullivan}
{Sullivan}, M., {Howell}, D.~A., {Perrett}, K., {et~al.} 2006, \aj, 131, 960,
  \dodoi{10.1086/499302}

\bibitem[{{The PLAsTiCC team} {et~al.}(2018){The PLAsTiCC team}, {Allam},
  {Bahmanyar}, {Biswas}, {Dai}, {Galbany}, {Hlo{\v{z}}ek}, {Ishida}, {Jha},
  {Jones}, {Kessler}, {Lochner}, {Mahabal}, {Malz}, {Mandel},
  {Mart{\'\i}nez-Galarza}, {McEwen}, {Muthukrishna}, {Narayan}, {Peiris},
  {Peters}, {Ponder}, {Setzer}, {The LSST Dark Energy Science Collaboration},
  {LSST Transients}, \& {Variable Stars Science Collaboration}}]{plasticc_data}
{The PLAsTiCC team}, {Allam}, Tarek, J., {Bahmanyar}, A., {et~al.} 2018, arXiv
  e-prints, arXiv:1810.00001.
\newblock \doarXiv{1810.00001}

\bibitem[{{Tolstikhin} {et~al.}(2021){Tolstikhin}, {Houlsby}, {Kolesnikov},
  {Beyer}, {Zhai}, {Unterthiner}, {Yung}, {Steiner}, {Keysers}, {Uszkoreit},
  {Lucic}, \& {Dosovitskiy}}]{MLPMixer}
{Tolstikhin}, I., {Houlsby}, N., {Kolesnikov}, A., {et~al.} 2021, arXiv
  e-prints, arXiv:2105.01601.
\newblock \doarXiv{2105.01601}

\bibitem[{{Villar} {et~al.}(2017{\natexlab{a}}){Villar}, {Berger}, {Metzger},
  \& {Guillochon}}]{SNII_4}
{Villar}, V.~A., {Berger}, E., {Metzger}, B.~D., \& {Guillochon}, J.
  2017{\natexlab{a}}, \\apj, 849, 70, \dodoi{10.3847/1538-4357/aa8fcb}

\bibitem[{{Villar} {et~al.}(2017{\natexlab{b}}){Villar}, {Berger}, {Metzger},
  \& {Guillochon}}]{SNIbc_4}
---. 2017{\natexlab{b}}, \\apj, 849, 70, \dodoi{10.3847/1538-4357/aa8fcb}

\bibitem[{{Villar} {et~al.}(2020{\natexlab{a}}){Villar}, {Cranmer}, {Contardo},
  {Ho}, \& {Yao-Yu Lin}}]{villar}
{Villar}, V.~A., {Cranmer}, M., {Contardo}, G., {Ho}, S., \& {Yao-Yu Lin}, J.
  2020{\natexlab{a}}, arXiv e-prints, arXiv:2010.11194.
\newblock \doarXiv{2010.11194}

\bibitem[{{Villar} {et~al.}(2020{\natexlab{b}}){Villar}, {Hosseinzadeh},
  {Berger}, {Ntampaka}, {Jones}, {Challis}, {Chornock}, {Drout}, {Foley},
  {Kirshner}, {Lunnan}, {Margutti}, {Milisavljevic}, {Sanders}, {Pan}, {Rest},
  {Scolnic}, {Magnier}, {Metcalfe}, {Wainscoat}, \& {Waters}}]{superraenn}
{Villar}, V.~A., {Hosseinzadeh}, G., {Berger}, E., {et~al.} 2020{\natexlab{b}},
  \apj, 905, 94, \dodoi{10.3847/1538-4357/abc6fd}

\bibitem[{{Woosley} {et~al.}(1987){Woosley}, {Pinto}, {Martin}, \&
  {Weaver}}]{SNIIb}
{Woosley}, S.~E., {Pinto}, P.~A., {Martin}, P.~G., \& {Weaver}, T.~A. 1987,
  \apj, 318, 664, \dodoi{10.1086/165402}

\bibitem[{Zeiler \& Fergus(2014)}]{zeiler2014visualizing}
Zeiler, M.~D., \& Fergus, R. 2014, in European conference on computer vision,
  Springer, 818--833

\end{thebibliography}
\end{document}